\documentclass[aps,prb,floatfix,twocolumn,footinbib,groupedaddress,showpacs]{revtex4-1}
\usepackage{amsmath,epsfig,amssymb,bm,amsfonts}  %%added 120522
\begin{document}

\title{
Emergent incommensurate correlations in the frustrated ferromagnetic spin-1 chains
}

\author{Hyeong Jun Lee and MooYoung Choi}
\affiliation{Department of Physics and Astronomy and Center for Theoretical Physics,
Seoul National University, Seoul 08826, Korea}
\author{Gun Sang Jeon}
\email{gsjeon@ewha.ac.kr}
\affiliation{Department of Physics, Ewha Womans University, Seoul 03760, Korea}

\date{\today}

\begin{abstract}    %%%%%%%%%%%%%%%%%%%%%%%%%%%%%%%%%%%%%%%%%%%%%%%%%%%%%
  We study the frustrated ferromagnetic spin-1 chains, where the ferromagnetic nearest-neighbor coupling competes with the antiferromagnetic next-nearest-neighbor coupling. 
  We use the density matrix renormalization group to obtain the ground states.
  Through the analysis of spin-spin correlations we identify the double Haldane phase as well as the ferromagnetic phase.
  It is shown that the ferromagnetic coupling leads to incommensurate correlations in the double Haldane phase. 
  Such short-range correlations transform continuously into the ferromagnetic instability at the transition to the ferromagnetic phase.
  We also compare the results with the spin-1/2 and classical spin systems, and discuss the string orders in the system.
\end{abstract}


\maketitle

\section{Introduction}    %%%%%%%%%%%%%%%%%%%%%%%%%%%%%%%%%%%%%%%%%%%%%%%%%%%%%%%%%

One-dimensional quantum spin systems have attracted
much interest for the past decade.
One of the reasons is that
quantum fluctuations play a more dominant role
in their ground states than for higher-dimensional systems.~\cite{Diep2005}
Generally quantum fluctuations suppress long-range order
of a many-body system in low dimensions.
	In the presence of antiferromagnetic nearest-neighbor (NN)
interactions the integer/half-integer Heisenberg spin chains  
were suggested to
have ground states with/without excitation gaps;~\cite{Haldane1983a,Haldane1983b}
this was confirmed by extensive numerical studies 
in the spin-1/2~\cite{White1993a,White1996}
and in the spin-1~\cite{Nightingale1986,Takahashi1989,Kennedy1990,White1992,White1993,White1993a,Golinelli1994} systems.
The resulting ground state of the spin-1 system 
is known to have a finite gap~\cite{White1992}
and a finite correlation length.~\cite{White1993,Schollwock1996}
It was also verified~\cite{Kennedy1990}
that the ground state is connected to the 
Affleck-Kennedy-Lieb-Tasaki (AKLT) state,~\cite{Affleck1987,Affleck1988}
which was originally obtained in the presence of bilinear and biquadratic interactions.
In addition to theoretical studies,
there have been extensive experimental studies:~\cite{Renard1988,Katsumata1989,Hagiwara1990,Glarum1991,Ma1992,Zaliznyak1994,Kenzelmann2001,Kenzelmann2002}
Some of the materials considered can be understood as one-dimensional spin-1 systems,
giving  supports for the theoretical results.
The data on Ni(C$_2$H$_8$N$_2$)$_2$NO$_2$ClO$_4$ 
provide evidence for the Haldane gap
\cite{Renard1988,Katsumata1989}
and additional spin-1/2 degrees of freedom 
\cite{Hagiwara1990,Glarum1991}
at the singlet-bond-broken sites
in the presence of impurities.
Gapped excitations were also observed in 
CsNiCl$_3$
with a correlation length smaller than
the prediction from the theory,~\cite{Zaliznyak1994,Kenzelmann2001,Kenzelmann2002}
which is now explained by weak couplings between spin-1 chains.
The zigzag spin chain in NaV(WO$_4)_2$ manifests a spin gap in the magnetic susceptibility.~\cite{Masuda2002}
Measurements of magnetic properties of 
ANi$_2$V$_2$O$_8$ (A=Pb,Sr) have also shown that they are described as a Haldane gapped phase without long-range order.~\cite{Uchiyama1999,Pahari2006,He2008}

Frustration, which is caused by 
the competition of two or more exchange interactions
or the system geometry, 
turns out to generate enormous quantum emergent phenomena 
in exotic phases.
Extensive numerical studies on frustrated spin chains 
have been performed.~\cite{Hikihara2001,White1996,Kolezhuk1996,Kolezhuk1997,Pixley2014}
One good example is 
the spin-1 chain with 
two types of antiferromagnetic interactions, one between 
	NN
spins and the other between next-nearest-neighbor (NNN) spins.~\cite{Kolezhuk1996,Kolezhuk1997,Pixley2014} 
The chain exhibits a discontinuous phase transition
between two kinds of the AKLT states;
they are distinguished by the patterns of singlet bonds in the chain.
The two phases are also characterized by
the existence of the hidden order due to breaking of the $Z_2 \times Z_2$ symmetry.

On the other hand, we have another kind of frustrated system, so-called frustrated ferromagnetic one, where 
the NN coupling is ferromagnetic.
Numerous studies have been performed on such frustrated ferromagnetic systems, especially for the spin-1/2 case.~\cite{Hamada1988,White1996,Allen1997,Hartel2008,Sudan2009,Sirker2010,Sirker2011,Furukawa2010,Furukawa2012}
It has been revealed that 
the spectrum data in cuprate materials such as LiCu$_2$O$_2$ and LiCuVO$_4$
can be explained as the properties of the frustrated ferromagnetic system
with good agreement.~\cite{Furukawa2010}
It is also found that interesting phase transitions exist at zero temperature
in this system.
This motivates us to examine a frustrated ferromagnetic spin-1 system,
of which the full understanding still lacks.

In this paper,
we study frustrated ferromagnetic quantum spin-1 systems
in one dimension.
We obtain the ground state 
through the use of the density-matrix-renormalization-group (DMRG) method, and
analyze the resulting spin-spin correlation functions.
It is found that 
incommensurate short-range correlations are induced in this system.
Such correlations turn out to be smoothly connected to the ferromagnetic phase at the transition with divergent correlation length.
We also discuss the string order parameters in the double Haldane phase.
This paper is organized as follows. In Sec. II, we describe the model and the method. Section III is devoted to the presentation and the discussion of numerical results.
Finally, a summary is given in Sec. IV.

\section{Model and Method}    %%%%%%%%%%%%%%%%%%%%%%%%%%%%%%%%%%%%%%%%%%%%%%%%%%%%%%%%%

We consider a one-dimensional spin-1 system with NN and NNN couplings.
The model is described by the Hamiltonian 
\begin{equation}
  \mathcal{H} = J_1 \sum_i \hat{\bm{S}}_i \cdot \hat{\bm{S}}_{i+1} +
  J_2 \sum_i \hat{\bm{S}}_i \cdot \hat{\bm{S}}_{i+2}
  \label{eq:H}
\end{equation}
where $\hat{\bm{S}}_i{\equiv}(\hat{S}^{x}_{i}, \hat{S}^{y}_{i}, \hat{S}^{z}_{i})$ is a vector spin-1 operator at the $i$th site.
The NN and the NNN exchange couplings 
are denoted by $J_1$ and $J_2$, respectively.
We investigate  
the Hamiltonian with ferromagnetic NN couplings ($J_1<0$)
and antiferromagnetic couplings ($J_2>0$) at zero temperature
in this work.

In the absence of the NN coupling ($J_1=0$) the system reduces to two decoupled subchains;
each consists of NNN pairs interacting with
the antiferromagnetic NNN interaction.
The individual subchain lies in the Haldane phase, which is characterized by
an excitation gap, exponentially decaying spin-spin correlations, and
long-range string order.~\cite{White1993}
Such a phase is called a double Haldane phase, in which singlet bonds are formed between NNN pairs.

In the presence of the antiferromagnetic NN coupling $(J_1>0)$
it was revealed~\cite{Kolezhuk1996,Kolezhuk1997}
that the double Haldane
phase undergoes a discontinuous transition to the Haldane phase
at $J_2/J_1 \simeq 0.744$.
At the transition the system exhibits a jump in the string order parameter with the bulk gap and the correlation length remaining finite.
The transition is attributed to 
the breaking of a hidden
$Z_2 \times Z_2$ symmetry in the Haldane phase.~\cite{Kennedy1992}
%	The disorder and the Lifshitz points was also identified in the Haldane phase as quantum remnants from the classical model.~\cite{Kolezhuk1996,Kolezhuk1997}

	When the ferromagnetic NN coupling is dominant ($|J_1| \gg J_2$),
the ground state is expected to be a ferromagnetically ordered state
with total spin $S_{\rm tot}=L$ with 
gapless spin-wave excitations.~\cite{[See {\em e.g.} ] Nomura1991}
It is also known that the ferromagnetic phase is robust
against additional competing interactions.
	In the intermediate region the thermodynamic behavior for the $S{=}1$ chains is less clear 
in contrast with spin-1/2 chain systems.~\cite{Hamada1988,White1996,Allen1997,Hartel2008,Sirker2010,Sirker2011}

\begin{figure}[t]
  \includegraphics[viewport=0 0 660 420,width=\linewidth]{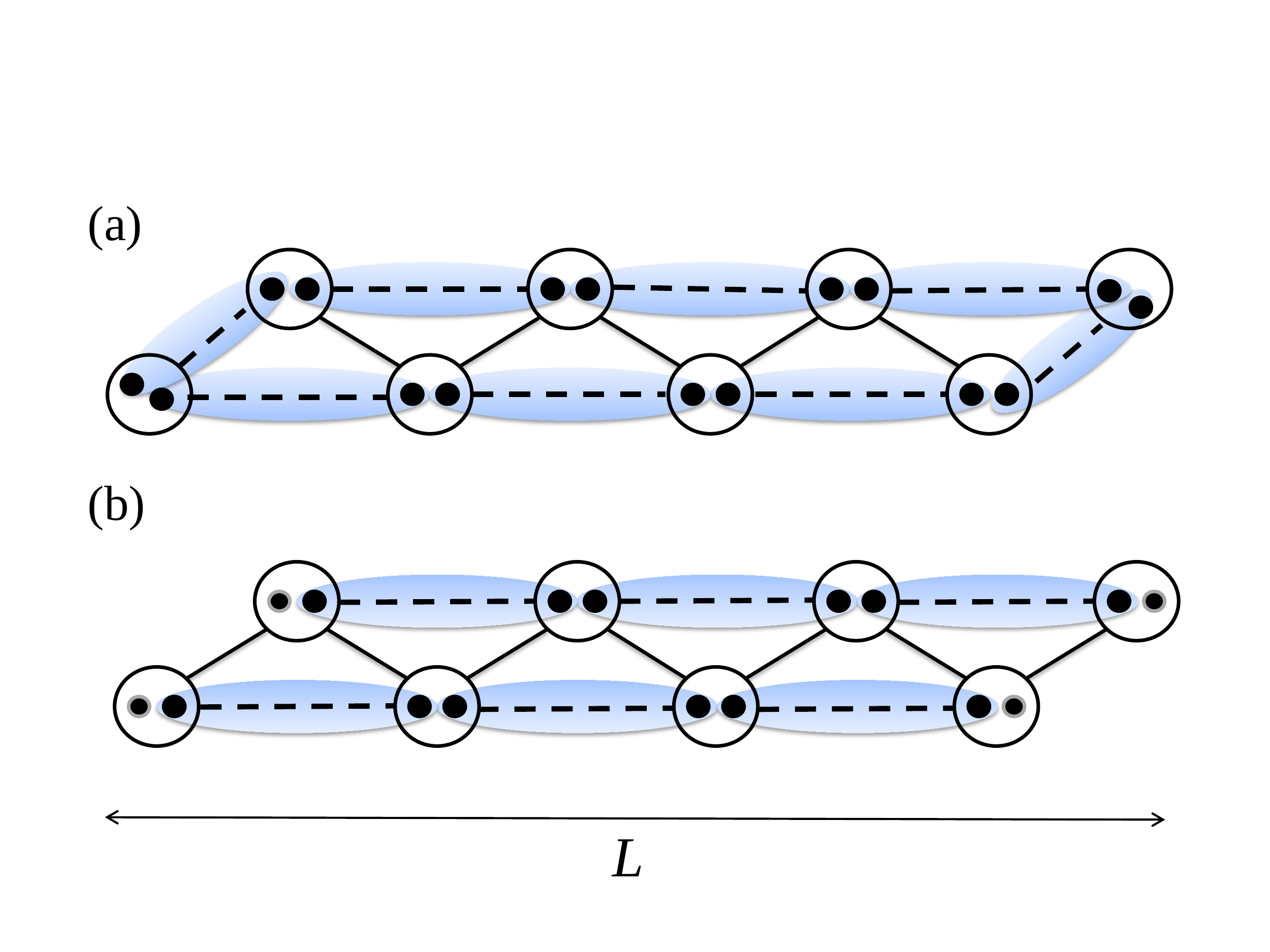}
  \caption{
  (Color online) 
  Schematic representation of a spin-1 chain in (a) modified and (b) standard open boundary conditions.
  The NN coupling $J_1$ and the NNN coupling $J_2$ are represented by solid and dashed lines, respectively.
  Shaded regions denote the singlet bonds which are expected to be formed between two half-spins of two sites in the double Haldane phase with $J_1<0$ and $J_2>0$.
  }
  \label{fig:rep}
\end{figure}

In this work,
we use the DMRG method~\cite{White1992,White1993,Schollwock2005}
with the infinite algorithm to obtain the ground state of the system. 
We adopt slightly modified open boundary conditions, which are sketched in Fig.~\ref{fig:rep}(a): the last two NN spins interact with $J_2$.
In the standard open boundary conditions in Fig.~\ref{fig:rep}(b)
the system tends to have two free 1/2-spins at each end, one on each subchain.
The antiferromagnetic interaction $J_2$ in the modified open boundary conditions helps the two free spins in standard open boundary conditions to form a singlet bond, leading to an effective finite-size calculation in the 
	double Haldane phase.
We also performed comparative computation in the standard open boundary conditions and found 
no significant qualitative difference in the bulk state. 

We have performed calculations
in chains up to $L=200$.
	The spin-spin correlations in different sizes turn out to collapse to a single curve in a wide range of coupling parameters as will be displayed later.
	We estimate the uncertainty in the data by the maximum deviation from the average value for the systems with size $L \ge 50$ and mark the error bars when they are larger than the size of symbols.
We have also checked out the convergence of the data with respect to $m$, 
where $m$ is the number of states per block after truncation,
by increasing $m$ consecutively.
The truncation errors in our calculations with $m=200$
are of the order of $10^{-5}$ to $10^{-7}$ for the 
	double Haldane
phase, in which the main interest of this work lies.
They are larger in the central region of the double Haldane
phase due to the increase of frustration.
In the region of the ferromagnetic phase the truncation errors reduce significantly below $10^{-15}$. 
We have performed the calculation with $m$ up to 250, and found that most physical results do not depend significantly on $m$.
Some $m$-sensitive physical quantities such as the correlation length have been presented together with the extrapolation value to $m=\infty$,
and relevant error bars have been marked when they are larger than the size of the symbols.
Henceforth
we will denote the energy and the length in units of $J_2$ 
and of the lattice constant, respectively, throughout this paper.

\section{Results and Discussion}    %%%%%%%%%%%%%%%%%%%%%%%%%%%%%%%%%%%%%%%%%%%%%%%%%%%%%%%%%
\begin{figure}[t]
  \includegraphics[width=0.9\linewidth]{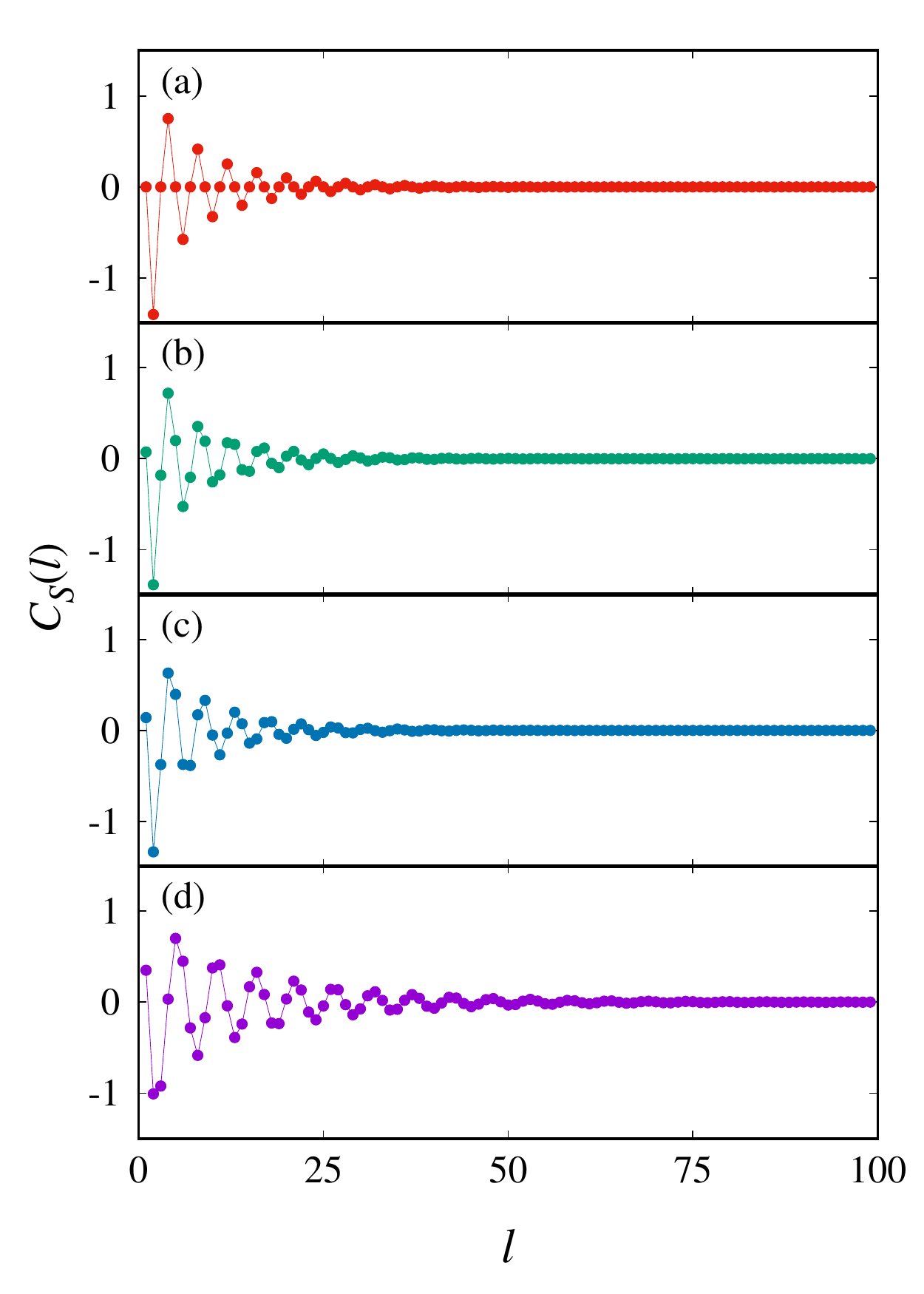}
  \caption{
  (Color online)
  Spin-spin correlation functions as a function of the separation $l$ between the spins
  at $J_1=0, -0.5, -1$, and $-2$ from top to bottom. 
  The period of the oscillation is 4 (i.e. $k=\pi/2)$ at $J_1=0$ and becomes incommensurate as $J_1$ decreases.
  }
  \label{fig:corr}
\end{figure}

To examine how the system 
evolves as the couplings vary,
we first calculate the spin-spin correlation function $C_S(l)$ 
defined by 
\begin{equation}
  C_S(l) \equiv \langle \hat{\bm{S}}_i \cdot \hat{\bm{S}}_{i+l} \rangle.
  \label{eq:corr}
\end{equation}
In order to minimize finite-size effects, we take site $i$ such that 
both sites $i$ and $i+l$ are as far from the boundaries as possible.
Figure~\ref{fig:corr}(a) shows the spin-spin correlation function
for $J_1=0$.
The ground state for $J_1=0$ can be
simply understood in terms of the two completely decoupled chains,
resulting in finite spin-spin correlations only for even-integer separations $l$ with vanishing correlations for odd-integer separations.
Within each subchain the system lies in the Haldane phase, which exhibits short-range antiferromagnetic correlations.
Such spin-spin correlations are in good agreement with those plotted in  
Fig.~\ref{fig:corr}(a). 
The correlations decay exponentially with the separation $l$, superposed by the oscillating correlations with the period of four.

We also plot the spin-spin correlations for various ferromagnetic NN couplings $J_1$ in Fig.~\ref{fig:corr}.
When $J_1$ is finite, the two subchains are no longer decoupled and finite correlations show up for odd-integer separations.
Interestingly, the correlations still display oscillating behavior apparently with a single period,
which is different from 4, the value for $J_1=0$;
 it increases monotonously with the increase of $|J_1|$.
It is also of interest to note that the oscillating period does not always appear commensurate with the lattice period.
Another conspicuous feature is that 
the decay of the correlations becomes slower as the ferromagnetic NN coupling becomes stronger.
While the correlations become negligible around $l\approx 30$ for $J_1=0$, 
we can observe clear oscillations for $l \gtrsim 60$ at $J_1=-2$. 
Until $J_1$ reaches the value $J_c \equiv -4$ the system does not show any abrupt change in the correlations, which implies that in the region of $J_c<J_1<0$ the system remains in the double Haldane phase.
For $J_1<J_c$, the spin-spin correlation function takes just the constant value of unity, independent of $l$, and the total spin of the ground state turns out to be $S_{\rm tot} = L$, signifying that the system is in the 
ferromagnetic phase. 
The critical value into the ferromagnetic phase for $S=1$ 
is the same as that in the case of classical spin systems~\cite{White1996} and $S=1/2$ spin systems,~\cite{Hamada1988,Hartel2008,Sudan2009,Sirker2010,Sirker2011} as pointed out in Ref.~\onlinecite{Bader1979}.

\begin{figure}[t]
  \includegraphics[width=\linewidth]{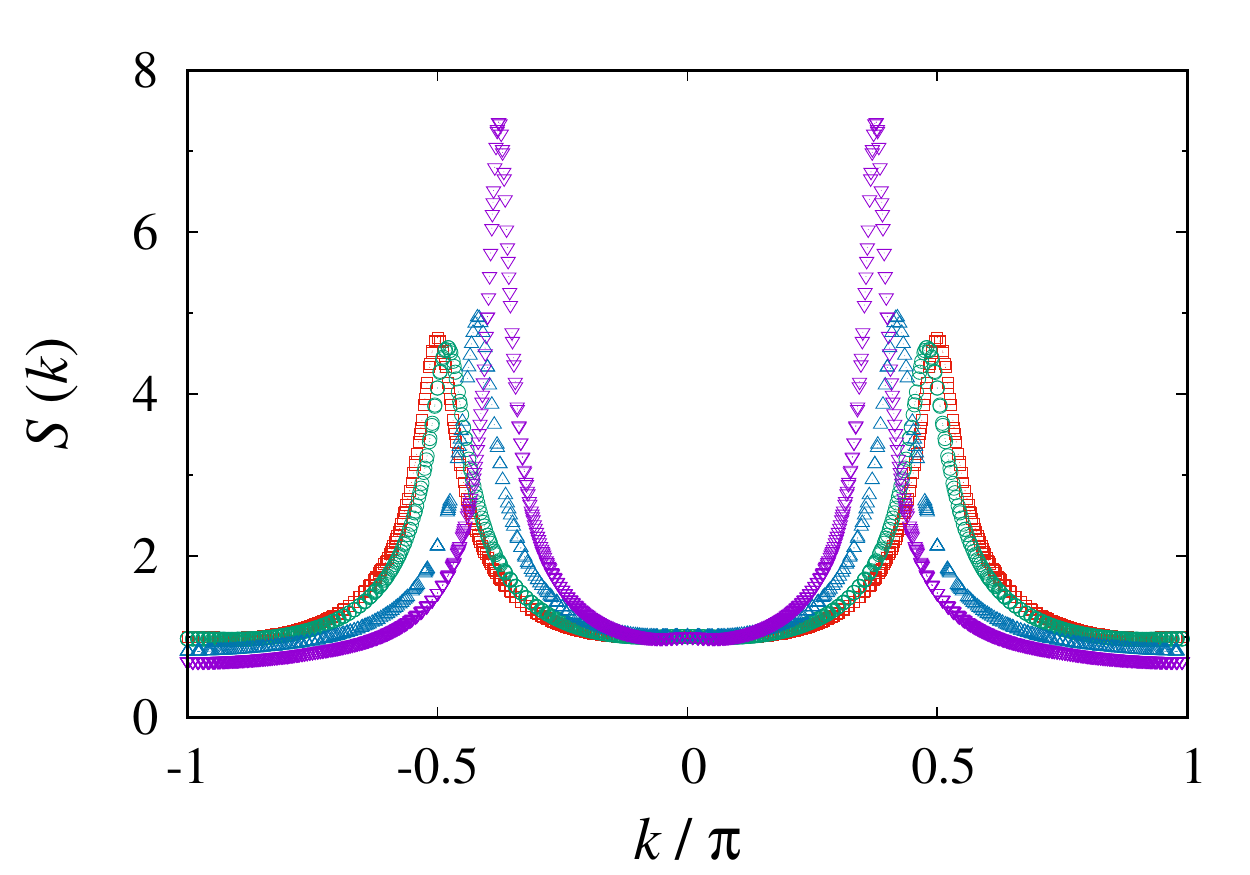}
  \caption{
  (Color online)
  Spin structure factor $S(k)$ for various NN couplings $J_1$.
  The data for $J_1=0$, $-0.5$, $-1$, and $-2$
  with various lengths $L=80$ to $200$
  are marked by (red) squares, (green) circles, (blue) triangles, and (purple) inverted triangles.
  The structure factor has two peaks in the 
	double Haldane phase, which are  
  not divergent in the thermodynamic limit.
  As $J_1$ is reduced, 
  the peaks move closer to $k=0$
  and the system exhibits incommensurate short-ranged correlations.
  }
  \label{fig:fft}
\end{figure}

For a quantitative analysis of the evolution of the states with $J_1$ varied, we examine the spin structure factor 
$$
S(k) \equiv \sum_l e^{ikl} C_S(l),
$$
which is the Fourier transform of $C_S(l)$.
We have used fast Fourier transform algorithms for $C_S(l)$ to obtain $S(k)$
and plot the resulting spin structure factor in Fig.~\ref{fig:fft} for various values of $J_1$.
For $J_1=0$ we have broad peaks at $k=\pm \pi/2$, which reflects oscillations with period 4.
The peak broadens due to the exponentially decaying correlations.
As $|J_1|$ increases, the positions of two peaks move towards $k=0$.
It is also clear that the peak gradually becomes narrower with the increase of $J_1$.
This result for $S(k)$ gives quantitative support to all the observations
on the spin-spin correlation function $C_S(l)$ mentioned above.
The structure factor remains finite even when $L$ is increased indefinitely, and the system does not exhibit long-range spin order in this region.

\begin{figure}[t]
  \includegraphics[width=\linewidth]{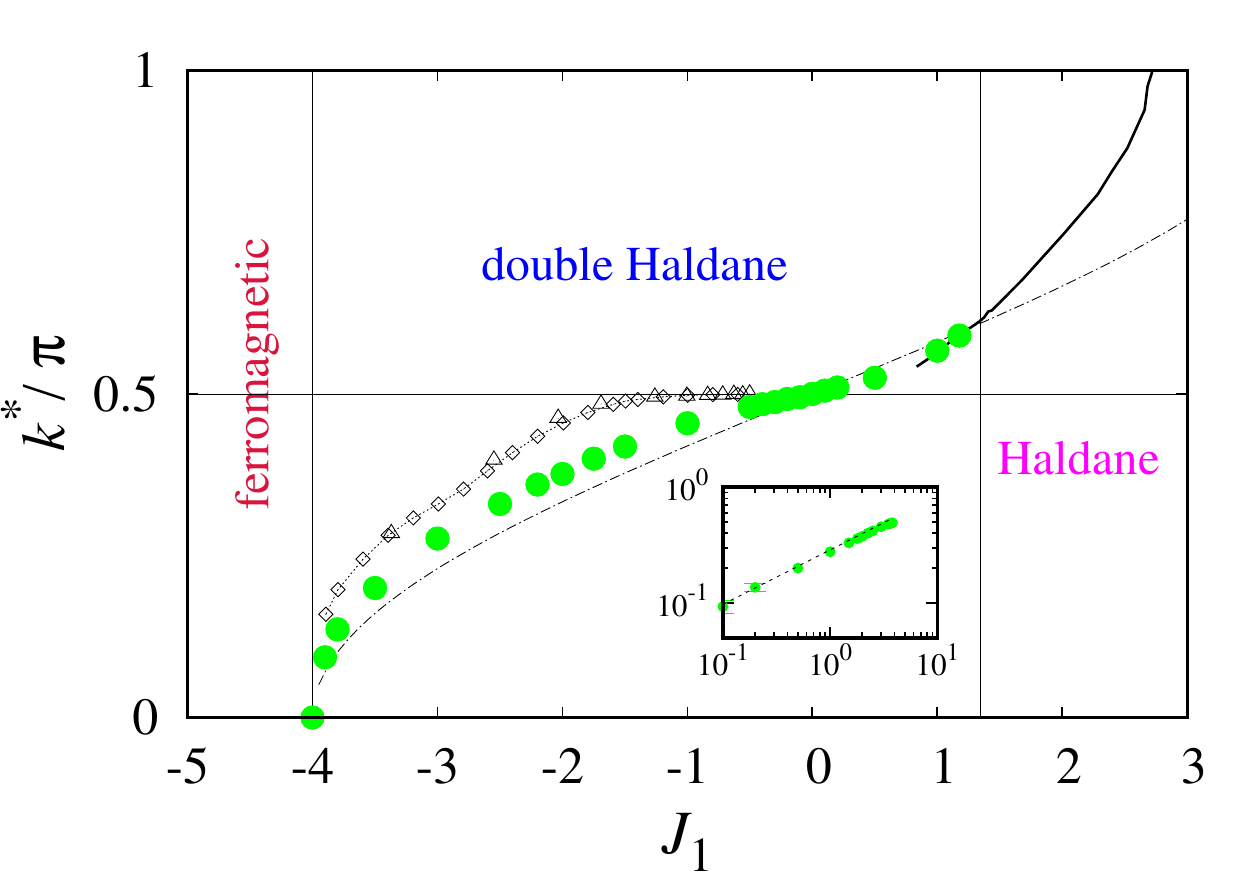}
  \caption{ (Color online)
  The pitch angle $k^*$ 
  as a function of the NN couplings $J_1$.
  The pitch angles, which are determined by the maximum position in the structure factor, are marked by (green) solid circles. 
  The pure NNN-AKLT state ($J_1=0$) has $k^*=\pi/2$
  reflecting the short-ranged antiferromagnetic correlation in each subchain.
 The antiferromagnetic correlation becomes
  incommensurate at finite $J_1$ in the 
	double Haldane phase. 
	  We also plot the results of the classical spin system, arccos$(-J_1/4)$ (dot-dashed line), and those of the spin-1/2 system from Ref.~\onlinecite{Sirker2010} (triangles) and Ref.~\onlinecite{Furukawa2012} (diamonds) for comparison.
  The data marked by a solid line as well as the transition point to the Haldane phase are quoted from Ref.~\onlinecite{Kolezhuk1997}.
  The inset shows the pitch angle in a log-log plot for $J_1-J_c >0$ 
  with $J_c=-4$ and the dashed line represents the best power-law fit to the data.
	  The data points without error bars have errors not larger than the size of the symbols.
  \label{fig:pospeak}
  }
\end{figure}

We can determine the pitch angle $k^*$ of the spin correlations by the position of the maximum in $S(k)$ on the side of positive $k$. 
For $J_1=0$, the peak is located at $k^*=\pi/2$, 
which is consistent with the NNN AKLT state, the prototype state in the double Haldane phase.
In Fig.~\ref{fig:pospeak} we plot $k^*$ as a function of $J_1$.
As $|J_1|$ increases, $k^*$ decreases monotonously from $k^*=\pi/2$ for $J_1=0$ to $k^*=0$ for $J_1=J_c$, and the ground state connects smoothly with the 
ferromagnetic state at $J_1=J_c$.
The decreasing curve corresponds to a convex-up function.

Such behavior is reminiscent of the spiral state which shows up in the classical spin system.
In the presence of NN ferromagnetic couplings and NNN antiferromagnetic couplings,
the classical spins exhibit a spiral state for $- J_c < J_1 < J_c$ with
the wave number given by $q= \textrm{arccos}(-J_1 / 4)$.~\cite{White1996}
Similar behaviors of the pitch angle were also reported in previous numerical studies of the spin-1/2 chains via the transfer-matrix DMRG method~\cite{Sirker2010} and the infinite time evolving block decimation algorithm method~\cite{Furukawa2010,Furukawa2012}.
For comparison, we have also plotted the data for classical and spin 1/2 chains with a dot-dashed line and empty symbols, respectively, in Fig.\ref{fig:pospeak}.
In all the three systems the pitch angle $k^*$ reduces with the increase of $|J_1|$ starting from $k^*=\pi/2$ at $J_1=0$, and approaches $k^*=0$ continuously at the transition into the ferromagnetic phase.
In the case of the spin-1/2 chains, the plateau-like region persists near $J_1=0$ up to $J_1\approx-2$.
On the other hand, the classical spin system displays rather gradual decrease even near $J_1=0$.
The curve of $k^*$ for the spin-1 chain locates between the classical and the spin-1/2 systems, which may be attributed to the reduction of quantum fluctuations in spin-1 systems in comparison with spin-1/2 systems.
It would be interesting to study variations of the pitch angle for higher spins, 
which should reflect the effects of both the changes in quantum fluctuations and the alternating behavior of integer and half-integer spins. 

In the inset of Fig.~\ref{fig:pospeak}, we plot $k^*$ as a function of $J_1-J_c$ in the log-log scale.
Near the critical point
the pitch angle displays the power-law behavior
\begin{equation}
 k^* \sim (J^{}_1 -J_c)^{\alpha}. 
  \label{eq:exp}
\end{equation}
The best fit in the range $-3.9 \leq J_1 \leq -3.0$ gives the exponent $\alpha=0.47(3)$.
Although
the best-fit value of $\alpha$ is a bit smaller than $\alpha_{\rm cl}=0.5$ for the classical spiral state, the two values are consistent within numerical errors. 
Numerical errors in the exponent are mainly due to the large relative errors near $J_c$.
It is also notable that the curve of $k^*$ versus $J_1-J_c$ is slightly convex up in the log-log plot, which tends to give an additional underestimate of $\alpha$ in the power-law fit over a finite window of $J_1$.

In Fig.~\ref{fig:pospeak} we also plot the pitch angle in the case of $J_1>0$: 
the solid line represents the data from an earlier study~\cite{Kolezhuk1997} for $S=1$
and the data points denoted by solid circles are obtained from our calculation.
The double Haldane phase in this region gives pitch angles in the range $\pi/2 < k^* < \pi$.~\cite{Kolezhuk1996,Kolezhuk1997}
The pitch angle increases with $J_1$ and becomes commensurate with $k^*=\pi$ around $J_1 \approx 2.67$.
Remarkably, the incommensurate spin correlations persist even in the region of the Haldane phase, which sets in at $J_1 \approx 1.34$.
This is in contrast to the fact that the pitch angle reduces to $k^*=0$ exactly at the transition to the ferromagnetic phase for $J_1<0$.

\begin{figure}[tp]
  \includegraphics[width=\linewidth]{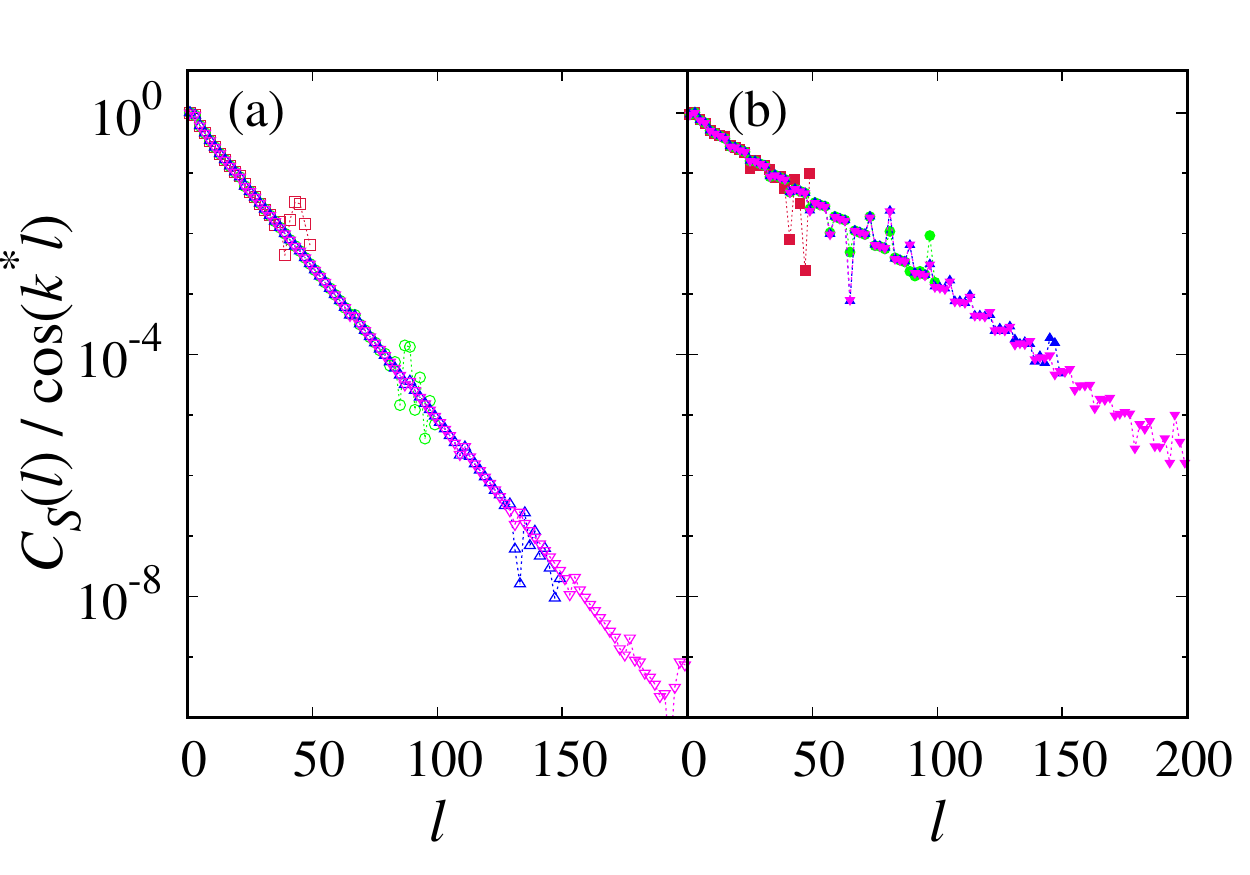}
  \caption{
  (Color online)
  Semi-log plot of the spin-spin correlation function $C_S(l)$ divided by $\cos(k^*l)$
  for several values of $L$ and (a) $J_1=-1$; (b) $J_1=-2$.
  The pitch angle $k^*$ is determined by the maximum position of the structure factor for each $J_1$.
  The data denoted by (red) squares, (green) circles, (blue) triangles, and (pink) inverted triangles correspond to $L=50, 100, 150$, and  $200$,  respectively.
  }
  \label{fig:diffL}
\end{figure}

\begin{figure}[tp]
  \includegraphics[width=\linewidth]{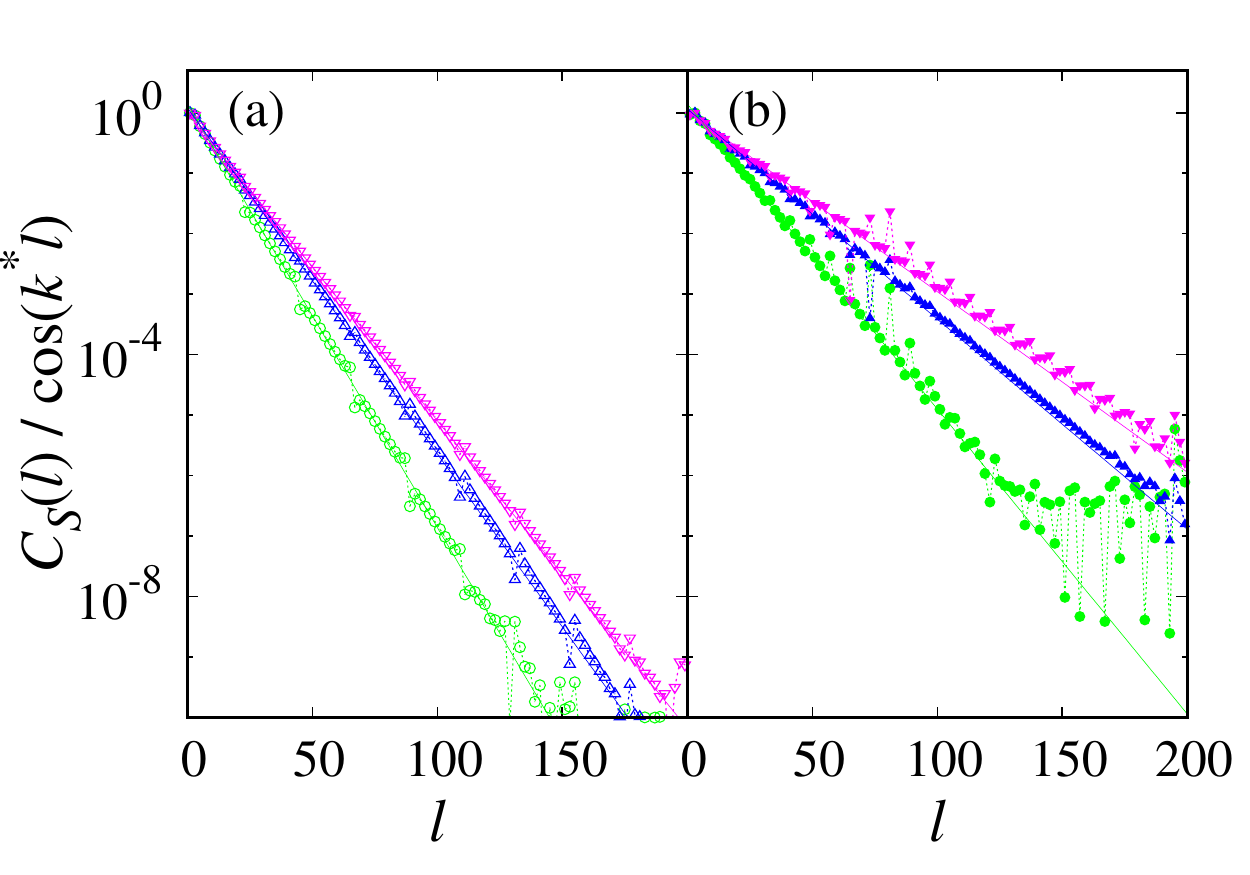}
  \caption{
  (Color online)
  Semi-log plot of the spin-spin correlation function $C_S(l)$ divided by $\cos(k^*l)$
   for several values of $m$ and (a) $J_1=-1$; (b) $J_1=-2$.
	   The pitch angle $k^*$ is determined in the same way as in Fig.~\ref{fig:diffL}.
  The data denoted by (green) circles, (blue) triangles, and (pink) inverted triangles correspond to $m=50, 100$, and  $150$,  respectively,
 and the solid lines are the best linear fits.
  }
  \label{fig:fit}
\end{figure}

\begin{figure}[tp]
  \includegraphics[width=\linewidth]{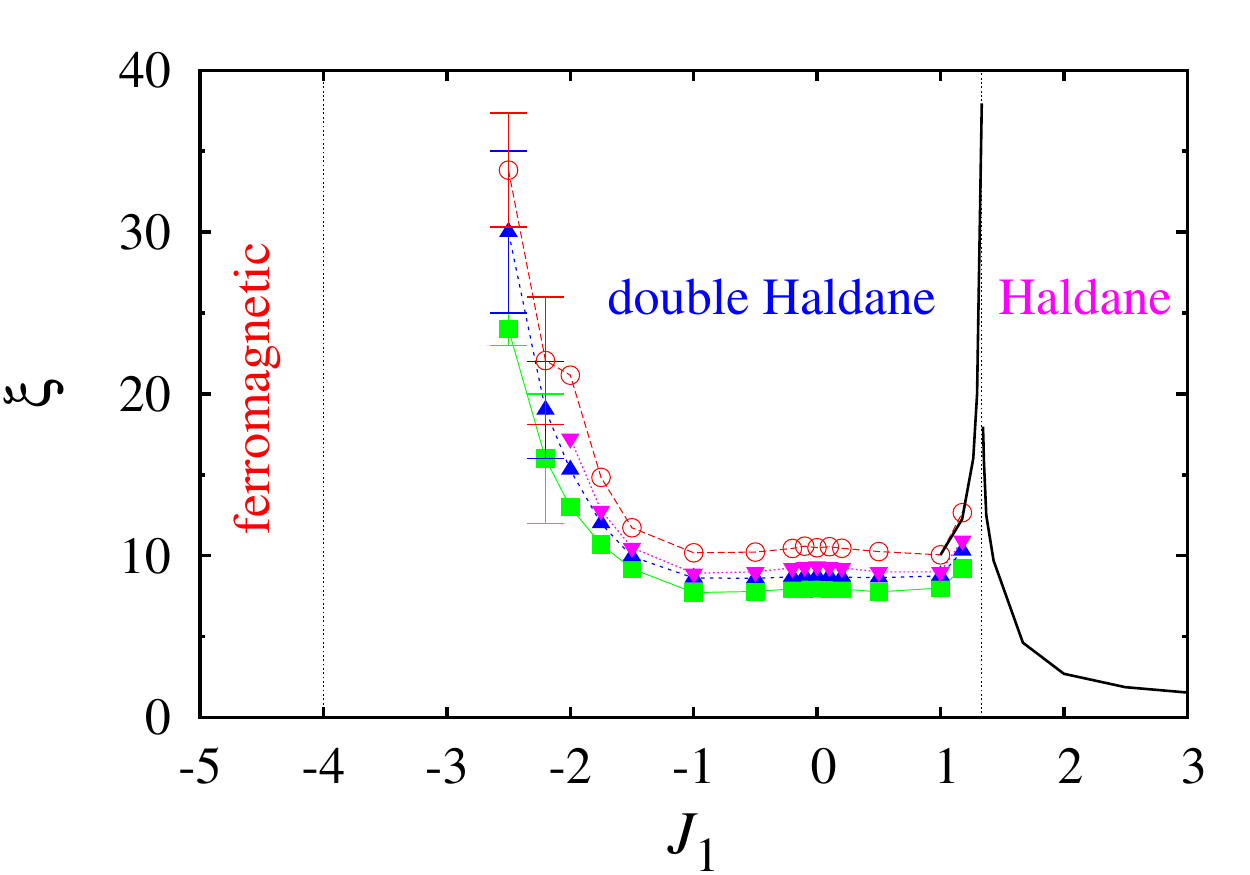}
  \caption{
  (Color online)
  Correlation length as a function of the NN coupling $J_1$
  for $m=$100 [(green) squares], 150 [(blue) triangles], 200 [(pink) inverted triangles], and its extrapolation to $m=\infty$ [(red) circles].
  As we approach the phase boundary to the ferromagnetic phase
  the correlation length becomes very large and is expected to diverge at $J=-4$.
  Solid lines denote the data quoted from Ref.~\onlinecite{Kolezhuk1996}.
	  The data points without error bars have errors not larger than the size of the symbols.
}
  \label{fig:xi}
\end{figure}

In order to examine the correlation length, 
we divide the spin correlation function  $C_S(l)$ by the oscillating factor $\cos(k^* l)$ and estimate the correlation length $\xi$ by fitting it to the exponential decay $\sim \exp(-l/\xi)$. 
	In Figs.~\ref{fig:diffL} and \ref{fig:fit}
one can see that $C_S(l)/\cos(k^* l)$ for $J=-1$ and $-2$ decays exponentially in a wide range of $l$.
	Figure~\ref{fig:diffL} also demonstrates that the quantities in different sizes exhibit almost the same exponentially-decaying behavior except for near the edges. 
The decay tends to become slower as the number $m$ of the kept states is increased.
In Fig.~\ref{fig:xi} we plot the correlation length $\xi$ estimated from the best fit as a function of $J_1$ for various values of $m$ as well as in the limit  $m \rightarrow \infty$.
We have also reproduced the data for $J_1>0$ from an earlier work.~\cite{Kolezhuk1996} 
For $J_1>0$ our numerical results are consistent with the peak associated with the transition between the Haldane and the double Haldane phases.
For $J_1<0$ the correlation length is enhanced as $J_1$ approaches $J_c$.
We also note that there exists a small bump around $J_1=0$, which signifies stronger spin fluctuations near the decoupled subchains.

\begin{figure}[tp]
  \includegraphics[width=\linewidth]{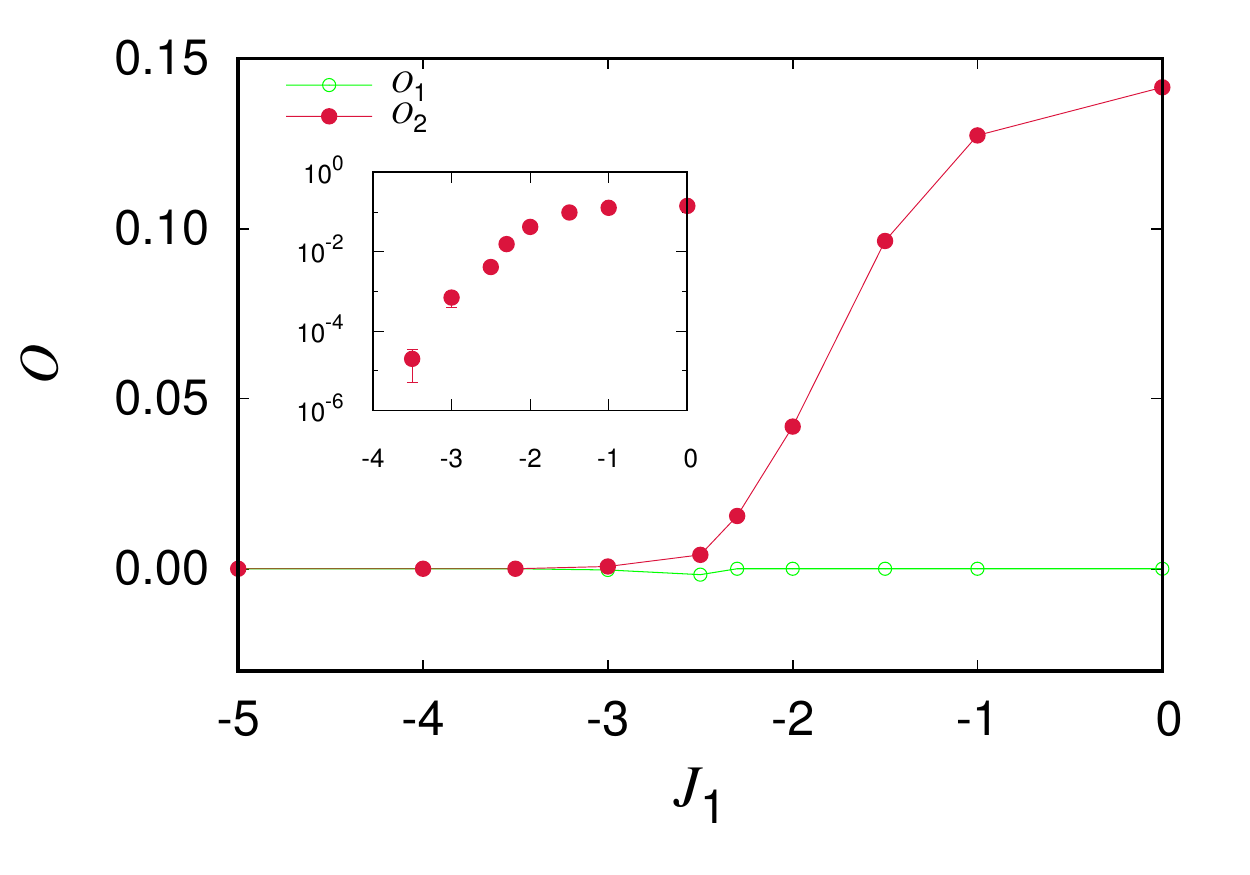}
  \caption{
  (Color online)
  String order parameters as a function of the NN coupling $J_1$.
  We represent $O_1$ and $O_2$ by (red) solid circles and (green) empty circles, respectively.
	    The inset displays the semi-log plot of $O_2$.
	  The data points without error bars have errors not larger than the size of the symbols.
}
  \label{fig:Ostr}
\end{figure}

We also examine 
string order and double-string order for $J_1<0$,  which can be probed by the use of the following nonlocal correlators~\cite{DenNijs1989,*Girvin1989,*Kennedy1992,Kolezhuk2002}
$$
O_1 (l,l') \equiv - \left\langle
S_l^z \left[
	\exp \left(\sum_{j{=}l+1}^{l'-1} i \pi S_j^z \right)
\right] S_{l'}^z
\right \rangle
$$
and
$$
O_2 (l,l') \equiv  \left\langle
S_l^z S_{l+1}^z \left[
	\exp \left(\sum_{j{=}l+2}^{l'-2} i \pi S_j^z \right)
\right] S_{l'-1}^z S_{l'}^z
\right \rangle.
$$
The string order parameters $O_1$ and $O_2$ can then be defined by
\begin{eqnarray}
	O_1 &\equiv& \lim_{|l-l'| \rightarrow \infty} O_1(l,l'),
\\
	O_2 &\equiv& \lim_{|l-l'| \rightarrow \infty} O_2(l,l')
\end{eqnarray}
in the limit of infinite separations.
Figure~\ref{fig:Ostr} shows string order parameters $O_1$ and $O_2$ versus $J_1$.
For $J_1=0$, we have nonzero $O_2$ with vanishing $O_1$, which is a typical characteristic of the double Haldane phase.
On the other hand, in the ferromagnetic phase ($J_1<J_c$) both $O_1$ and $O_2$ vanish.
As $J_1$ decreases from zero, $O_2$ also reduces.

It is remarkable that $O_2$ is significantly reduced for $J_1 \lesssim -2$ while the transition into the ferromagnetic phase occurs at $J_1=-4$.
Such an abrupt reduction may suggest the possible existence of emergent phase between the double Haldane phase and the ferromagnetic phase.
However, the detailed analysis has revealed that the string nonlocal correlator $O_2(l,l')$ shows quite distinct behavior for $-4 < J_1 \lesssim -2$ from that of the ferromagnetic phase where the string order is absent. 
In the ferromagnetic phase $O_2(l,l')$ shows a clear exponential decay with $|l-l'|$, implying that the corresponding string order parameter $O_2$ vanishes in the thermodynamic limit. 
For $-4 <J_1 \lesssim -2$, in contrast, $O_2(l,l')$ remains finite although its values are very small.
The semi-log plot of $O_2$, which is presented in the inset of Fig.~\ref{fig:Ostr}, also supports a direct transition from the double Haldane phase to a ferromagnetic one.
It demonstrates that
the decreasing behavior is rather consistent with the transition at $J_1=-4$
although a rapid decrease of $O_2$ starts around $J_1 \approx -2$.

\begin{figure}[tp]
  \includegraphics[width=\linewidth]{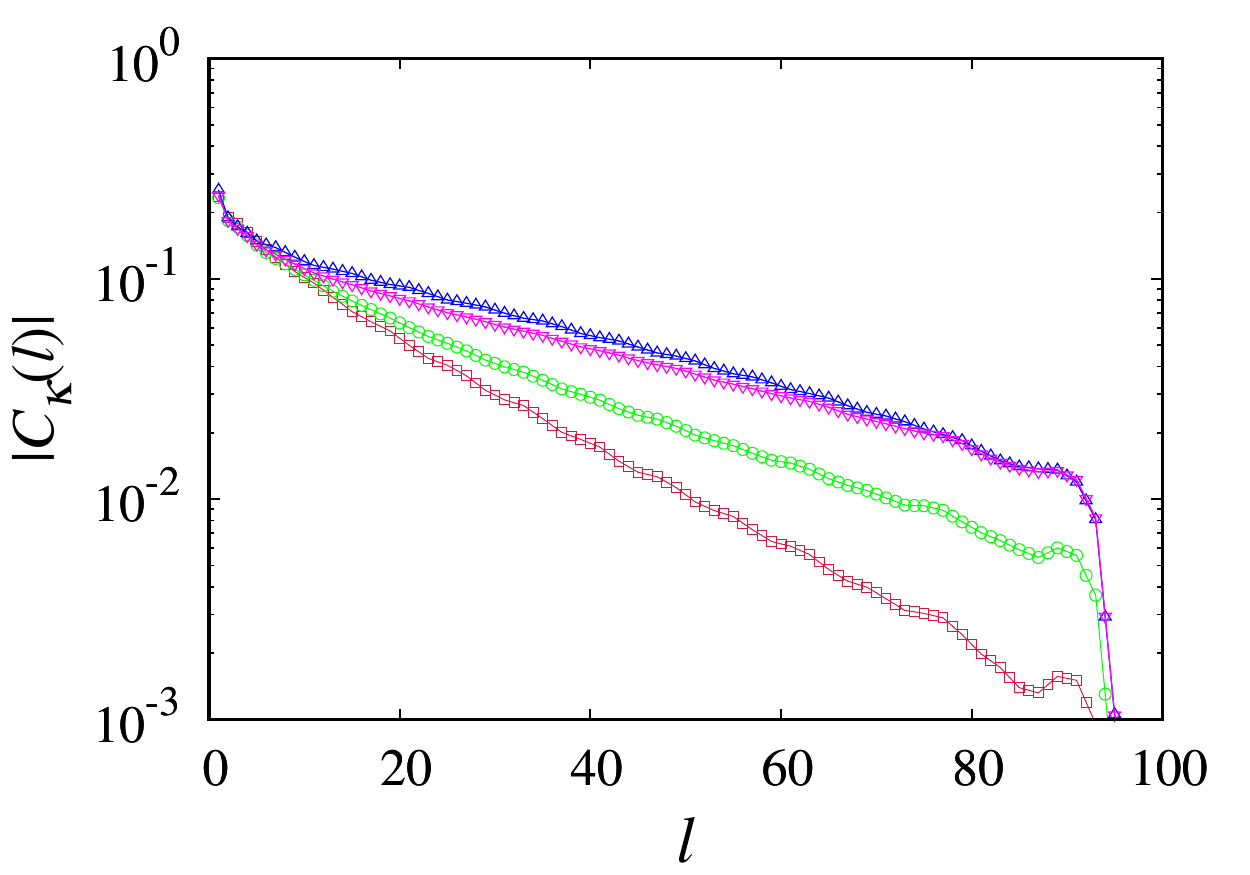}
  \caption{
  (Color online)
  Semi-log plot of the chirality correlation function $C_\kappa(l)$ 
  for $J_1=-3$.
  The data denoted by (red) squares, (green) circles, (blue) triangles, and (pink) inverted triangles correspond to $m=50, 100, 150$, and  $200$,  respectively.
  }
  \label{fig:Chi_corr}
\end{figure}

	 It is also worth while to examine whether a chiral phase is present between the double Haldane phase and the ferromagnetic phase.
	 In the spin-1/2 case the chiral phase has been reported in a close vicinity of the region,~\cite{Furukawa2010,Furukawa2012} and accordingly it is a strong candidate for a new phase if it emerges also in the spin-1 case.
We have thus computed the chirality correlation function defined by
\begin{equation}
	C_\kappa (l) \equiv \langle \hat{\kappa}_i^z \hat{\kappa}_{i+l}^z \rangle
\end{equation}
with 
\begin{equation}
	\hat{\kappa}_i^z \equiv \hat{z} \cdot (\hat{\bm{S}}_i \times \hat{\bm{S}}_{i+1} ) = \hat{S}^x_i \hat{S}^y_{i+1} - \hat{S}^y_i \hat{S}^x_{i+1}.
\end{equation}
Figure~\ref{fig:Chi_corr} presents $C_\kappa(l)$ for $J_1=-3$, where $O_2$ is significantly reduced.
The chirality correlation function exhibits an apparent exponential decrease, which indicates the absence of a chiral phase.
This also implies that the emergence of a new phase is less probable.
We presume that the great reduction in $O_2$ for $-4<J_1<-2$ is caused by the large ferromagnetic fluctuations which are enhanced markedly in that region as can be seen in Fig.~\ref{fig:xi}. 
\section{Summary}    %%%%%%%%%%%%%%%%%%%%%%%%%%%%%%%%%%%%%%%%%%%%%%%%%%%%%%%%%
We have investigated the one-dimensional spin-1 system frustrated
by the combination of the spin exchange interactions:
the ferromagnetic one between NN spins
and the antiferromagnetic one between NNN spins.
	Via the DMRG calculations we have confirmed explicitly 
that the ferromagnetic phase transition for $S=1$
occurs at $J_1=-4 (\equiv J_c)$,~\cite{Bader1979,Balents2016}
below which 
the spin-spin correlation function becomes constant. 
	The robustness of $J_c$ suggests that quantum fluctuations due to the quantum nature of the spin do not affect significantly the ferromagnetic phase transition; this is in sharp contrast with the fact that other phase boundaries of the frustrated spin chains exhibit strong dependence on $S$ and a variety of distinct phases.~\cite{Hikihara2001} 
Such interesting results may be tested	through experiments on ultracold atoms.~\cite{Senko2015,Cohen2015,Garcia-Ripoll2004}

In the double Haldane phase, on the other hand,
the system shows short-ranged antiferromagnetic spin-spin correlations.
Such behavior remains robust in a wide range of the NN ferromagnetic interaction strength $J_1$ up to $J_c$.
The pitch angle of the incommensurate spin-spin correlations decreases 
	from $\pi/2$
as $J_1$ approaches $J_c$, vanishing at $J_1=J_c$.
Similar behaviors were also reported in earlier works on the classical and spin-1/2 systems.
It has been revealed that the results of the spin-1 system are closer to the classical results than those of the spin-1/2 one.
We have also discussed the behavior of string order parameters in the double Haldane phase in the presence of ferromagnetic NN couplings.
It has been found that the string order parameter $O_2$ undergoes a substantial reduction far above $J_c$.
However, detailed analysis has suggested that a new phase is less likely to emerge in that region.
It is presumed that the enhancement in the ferromagnetic fluctuations gives rise to such substantial reduction in $O_2$.

	Various extensions of our study are expected to produce fruitful results in future study. 
	The introduction of the anisotropic interaction of the XXZ-type~\cite{Hikihara2001} in our model will reduce the rotational symmetry and possibly give rise to another symmetry-broken state such as the chiral or the dimer-like ones.
	It will deepen our understanding of the roles of quantum fluctuations in connection with the Kosterlitz-Thouless transition. 
It is also worth while to investigate the effects of the uniaxial single-ion anisotropy.~\cite{Hikihara2002}
It is expected to enrich the physics in quantum spin-1 chains and to draw out a variety of implications on experimental results.

\acknowledgments
This work
was supported by the National Research Foundation of Korea
through Grant No. 2008-0061893 (G.S.J.), 
Grant No. 2013R1A1A2007959 (H.J.L. and G.S.J.), 
and Grant No. 2016R1D1A1A09917318 (M.Y.C.).

\bibliography{QSC1.bib}

\begin{thebibliography}{50}%
\makeatletter
\providecommand \@ifxundefined [1]{%
 \@ifx{#1\undefined}
}%
\providecommand \@ifnum [1]{%
 \ifnum #1\expandafter \@firstoftwo
 \else \expandafter \@secondoftwo
 \fi
}%
\providecommand \@ifx [1]{%
 \ifx #1\expandafter \@firstoftwo
 \else \expandafter \@secondoftwo
 \fi
}%
\providecommand \natexlab [1]{#1}%
\providecommand \enquote  [1]{``#1''}%
\providecommand \bibnamefont  [1]{#1}%
\providecommand \bibfnamefont [1]{#1}%
\providecommand \citenamefont [1]{#1}%
\providecommand \href@noop [0]{\@secondoftwo}%
\providecommand \href [0]{\begingroup \@sanitize@url \@href}%
\providecommand \@href[1]{\@@startlink{#1}\@@href}%
\providecommand \@@href[1]{\endgroup#1\@@endlink}%
\providecommand \@sanitize@url [0]{\catcode `\\12\catcode `\$12\catcode
  `\&12\catcode `\#12\catcode `\^12\catcode `\_12\catcode `\%12\relax}%
\providecommand \@@startlink[1]{}%
\providecommand \@@endlink[0]{}%
\providecommand \url  [0]{\begingroup\@sanitize@url \@url }%
\providecommand \@url [1]{\endgroup\@href {#1}{\urlprefix }}%
\providecommand \urlprefix  [0]{URL }%
\providecommand \Eprint [0]{\href }%
\providecommand \doibase [0]{http://dx.doi.org/}%
\providecommand \selectlanguage [0]{\@gobble}%
\providecommand \bibinfo  [0]{\@secondoftwo}%
\providecommand \bibfield  [0]{\@secondoftwo}%
\providecommand \translation [1]{[#1]}%
\providecommand \BibitemOpen [0]{}%
\providecommand \bibitemStop [0]{}%
\providecommand \bibitemNoStop [0]{.\EOS\space}%
\providecommand \EOS [0]{\spacefactor3000\relax}%
\providecommand \BibitemShut  [1]{\csname bibitem#1\endcsname}%
\let\auto@bib@innerbib\@empty
%</preamble>
\bibitem [{\citenamefont {Diep}(2013)}]{Diep2005}%
  \BibitemOpen
  \bibinfo {editor} {\bibfnamefont {H.~T.}\ \bibnamefont {Diep}},\ ed.,\
  \href@noop {} {\emph {\bibinfo {title} {{Frustrated Spin Systems}}}},\
  \bibinfo {edition} {2nd}\ ed.\ (\bibinfo  {publisher} {World Scientific},\
  \bibinfo {year} {2013})\BibitemShut {NoStop}%
\bibitem [{\citenamefont {Haldane}(1983{\natexlab{a}})}]{Haldane1983a}%
  \BibitemOpen
  \bibfield  {author} {\bibinfo {author} {\bibfnamefont {F.~D.~M.}\
  \bibnamefont {Haldane}},\ }\href@noop {} {\bibfield  {journal} {\bibinfo
  {journal} {Phys. Rev. Lett.}\ }\textbf {\bibinfo {volume} {50}},\ \bibinfo
  {pages} {1153} (\bibinfo {year} {1983}{\natexlab{a}})}\BibitemShut {NoStop}%
\bibitem [{\citenamefont {Haldane}(1983{\natexlab{b}})}]{Haldane1983b}%
  \BibitemOpen
  \bibfield  {author} {\bibinfo {author} {\bibfnamefont {F.~D.~M.}\
  \bibnamefont {Haldane}},\ }\href@noop {} {\bibfield  {journal} {\bibinfo
  {journal} {Phys. Lett. A}\ }\textbf {\bibinfo {volume} {93}},\ \bibinfo
  {pages} {464} (\bibinfo {year} {1983}{\natexlab{b}})}\BibitemShut {NoStop}%
\bibitem [{\citenamefont {White}(1993)}]{White1993a}%
  \BibitemOpen
  \bibfield  {author} {\bibinfo {author} {\bibfnamefont {S.~R.}\ \bibnamefont
  {White}},\ }\href@noop {} {\bibfield  {journal} {\bibinfo  {journal} {Phys.
  Rev. B}\ }\textbf {\bibinfo {volume} {48}},\ \bibinfo {pages} {10345}
  (\bibinfo {year} {1993})}\BibitemShut {NoStop}%
\bibitem [{\citenamefont {White}\ and\ \citenamefont
  {Affleck}(1996)}]{White1996}%
  \BibitemOpen
  \bibfield  {author} {\bibinfo {author} {\bibfnamefont {S.~R.}\ \bibnamefont
  {White}}\ and\ \bibinfo {author} {\bibfnamefont {I.}~\bibnamefont
  {Affleck}},\ }\href@noop {} {\bibfield  {journal} {\bibinfo  {journal} {Phys.
  Rev. B}\ }\textbf {\bibinfo {volume} {54}},\ \bibinfo {pages} {9862}
  (\bibinfo {year} {1996})}\BibitemShut {NoStop}%
\bibitem [{\citenamefont {Nightingale}\ and\ \citenamefont
  {Bl\"ote}(1986)}]{Nightingale1986}%
  \BibitemOpen
  \bibfield  {author} {\bibinfo {author} {\bibfnamefont {M.~P.}\ \bibnamefont
  {Nightingale}}\ and\ \bibinfo {author} {\bibfnamefont {H.~W.~J.}\
  \bibnamefont {Bl\"ote}},\ }\href@noop {} {\bibfield  {journal} {\bibinfo
  {journal} {Phys. Rev. B}\ }\textbf {\bibinfo {volume} {33}},\ \bibinfo
  {pages} {659} (\bibinfo {year} {1986})}\BibitemShut {NoStop}%
\bibitem [{\citenamefont {Takahashi}(1989)}]{Takahashi1989}%
  \BibitemOpen
  \bibfield  {author} {\bibinfo {author} {\bibfnamefont {M.}~\bibnamefont
  {Takahashi}},\ }\href@noop {} {\bibfield  {journal} {\bibinfo  {journal}
  {Phys. Rev. Lett.}\ }\textbf {\bibinfo {volume} {62}},\ \bibinfo {pages}
  {2313} (\bibinfo {year} {1989})}\BibitemShut {NoStop}%
\bibitem [{\citenamefont {Kennedy}(1990)}]{Kennedy1990}%
  \BibitemOpen
  \bibfield  {author} {\bibinfo {author} {\bibfnamefont {T.}~\bibnamefont
  {Kennedy}},\ }\href@noop {} {\bibfield  {journal} {\bibinfo  {journal} {J.
  Phys.: Condens. Matter}\ }\textbf {\bibinfo {volume} {2}},\ \bibinfo {pages}
  {5737} (\bibinfo {year} {1990})}\BibitemShut {NoStop}%
\bibitem [{\citenamefont {White}(1992)}]{White1992}%
  \BibitemOpen
  \bibfield  {author} {\bibinfo {author} {\bibfnamefont {S.~R.}\ \bibnamefont
  {White}},\ }\href@noop {} {\bibfield  {journal} {\bibinfo  {journal} {Phys.
  Rev. Lett.}\ }\textbf {\bibinfo {volume} {69}},\ \bibinfo {pages} {2863}
  (\bibinfo {year} {1992})}\BibitemShut {NoStop}%
\bibitem [{\citenamefont {White}\ and\ \citenamefont {Huse}(1993)}]{White1993}%
  \BibitemOpen
  \bibfield  {author} {\bibinfo {author} {\bibfnamefont {S.~R.}\ \bibnamefont
  {White}}\ and\ \bibinfo {author} {\bibfnamefont {D.~A.}\ \bibnamefont
  {Huse}},\ }\href@noop {} {\bibfield  {journal} {\bibinfo  {journal} {Phys.
  Rev. B}\ }\textbf {\bibinfo {volume} {48}},\ \bibinfo {pages} {3844}
  (\bibinfo {year} {1993})}\BibitemShut {NoStop}%
\bibitem [{\citenamefont {Golinelli}\ \emph {et~al.}(1994)\citenamefont
  {Golinelli}, \citenamefont {Jolic\oe{}ur},\ and\ \citenamefont
  {Lacaze}}]{Golinelli1994}%
  \BibitemOpen
  \bibfield  {author} {\bibinfo {author} {\bibfnamefont {O.}~\bibnamefont
  {Golinelli}}, \bibinfo {author} {\bibfnamefont {T.}~\bibnamefont
  {Jolic\oe{}ur}}, \ and\ \bibinfo {author} {\bibfnamefont {R.}~\bibnamefont
  {Lacaze}},\ }\href@noop {} {\bibfield  {journal} {\bibinfo  {journal} {Phys.
  Rev. B}\ }\textbf {\bibinfo {volume} {50}},\ \bibinfo {pages} {3037}
  (\bibinfo {year} {1994})}\BibitemShut {NoStop}%
\bibitem [{\citenamefont {Schollw\"ock}\ \emph {et~al.}(1996)\citenamefont
  {Schollw\"ock}, \citenamefont {Jolic\oe{}ur},\ and\ \citenamefont
  {Garel}}]{Schollwock1996}%
  \BibitemOpen
  \bibfield  {author} {\bibinfo {author} {\bibfnamefont {U.}~\bibnamefont
  {Schollw\"ock}}, \bibinfo {author} {\bibfnamefont {T.}~\bibnamefont
  {Jolic\oe{}ur}}, \ and\ \bibinfo {author} {\bibfnamefont {T.}~\bibnamefont
  {Garel}},\ }\href@noop {} {\bibfield  {journal} {\bibinfo  {journal} {Phys.
  Rev. B}\ }\textbf {\bibinfo {volume} {53}},\ \bibinfo {pages} {3304}
  (\bibinfo {year} {1996})}\BibitemShut {NoStop}%
\bibitem [{\citenamefont {Affleck}\ \emph {et~al.}(1987)\citenamefont
  {Affleck}, \citenamefont {Kennedy}, \citenamefont {Lieb},\ and\ \citenamefont
  {Tasaki}}]{Affleck1987}%
  \BibitemOpen
  \bibfield  {author} {\bibinfo {author} {\bibfnamefont {I.}~\bibnamefont
  {Affleck}}, \bibinfo {author} {\bibfnamefont {T.}~\bibnamefont {Kennedy}},
  \bibinfo {author} {\bibfnamefont {E.~H.}\ \bibnamefont {Lieb}}, \ and\
  \bibinfo {author} {\bibfnamefont {H.}~\bibnamefont {Tasaki}},\ }\href@noop {}
  {\bibfield  {journal} {\bibinfo  {journal} {Phys. Rev. Lett.}\ }\textbf
  {\bibinfo {volume} {59}},\ \bibinfo {pages} {799} (\bibinfo {year}
  {1987})}\BibitemShut {NoStop}%
\bibitem [{\citenamefont {Affleck}\ \emph {et~al.}(1988)\citenamefont
  {Affleck}, \citenamefont {Kennedy}, \citenamefont {Lieb},\ and\ \citenamefont
  {Tasaki}}]{Affleck1988}%
  \BibitemOpen
  \bibfield  {author} {\bibinfo {author} {\bibfnamefont {I.}~\bibnamefont
  {Affleck}}, \bibinfo {author} {\bibfnamefont {T.}~\bibnamefont {Kennedy}},
  \bibinfo {author} {\bibfnamefont {E.~H.}\ \bibnamefont {Lieb}}, \ and\
  \bibinfo {author} {\bibfnamefont {H.}~\bibnamefont {Tasaki}},\ }\href@noop {}
  {\bibfield  {journal} {\bibinfo  {journal} {Commun. Math. Phys.}\ }\textbf
  {\bibinfo {volume} {115}},\ \bibinfo {pages} {477} (\bibinfo {year}
  {1988})}\BibitemShut {NoStop}%
\bibitem [{\citenamefont {Renard}\ \emph {et~al.}(1988)\citenamefont {Renard},
  \citenamefont {Verdaguer}, \citenamefont {Regnault}, \citenamefont
  {Erkelens}, \citenamefont {Rossat-Mignod}, \citenamefont {Ribas},
  \citenamefont {Stirling},\ and\ \citenamefont {Vettier}}]{Renard1988}%
  \BibitemOpen
  \bibfield  {author} {\bibinfo {author} {\bibfnamefont {J.~P.}\ \bibnamefont
  {Renard}}, \bibinfo {author} {\bibfnamefont {M.}~\bibnamefont {Verdaguer}},
  \bibinfo {author} {\bibfnamefont {L.~P.}\ \bibnamefont {Regnault}}, \bibinfo
  {author} {\bibfnamefont {W.~A.~C.}\ \bibnamefont {Erkelens}}, \bibinfo
  {author} {\bibfnamefont {J.}~\bibnamefont {Rossat-Mignod}}, \bibinfo {author}
  {\bibfnamefont {J.}~\bibnamefont {Ribas}}, \bibinfo {author} {\bibfnamefont
  {W.~G.}\ \bibnamefont {Stirling}}, \ and\ \bibinfo {author} {\bibfnamefont
  {C.}~\bibnamefont {Vettier}},\ }\href@noop {} {\bibfield  {journal} {\bibinfo
   {journal} {J. Appl. Phys.}\ }\textbf {\bibinfo {volume} {63}},\ \bibinfo
  {pages} {3538} (\bibinfo {year} {1988})}\BibitemShut {NoStop}%
\bibitem [{\citenamefont {Katsumata}\ \emph {et~al.}(1989)\citenamefont
  {Katsumata}, \citenamefont {Hori}, \citenamefont {Takeuchi}, \citenamefont
  {Date}, \citenamefont {Yamagishi},\ and\ \citenamefont
  {Renard}}]{Katsumata1989}%
  \BibitemOpen
  \bibfield  {author} {\bibinfo {author} {\bibfnamefont {K.}~\bibnamefont
  {Katsumata}}, \bibinfo {author} {\bibfnamefont {H.}~\bibnamefont {Hori}},
  \bibinfo {author} {\bibfnamefont {T.}~\bibnamefont {Takeuchi}}, \bibinfo
  {author} {\bibfnamefont {M.}~\bibnamefont {Date}}, \bibinfo {author}
  {\bibfnamefont {A.}~\bibnamefont {Yamagishi}}, \ and\ \bibinfo {author}
  {\bibfnamefont {J.~P.}\ \bibnamefont {Renard}},\ }\href@noop {} {\bibfield
  {journal} {\bibinfo  {journal} {Phys. Rev. Lett.}\ }\textbf {\bibinfo
  {volume} {63}},\ \bibinfo {pages} {86} (\bibinfo {year} {1989})}\BibitemShut
  {NoStop}%
\bibitem [{\citenamefont {Hagiwara}\ \emph {et~al.}(1990)\citenamefont
  {Hagiwara}, \citenamefont {Katsumata}, \citenamefont {Affleck}, \citenamefont
  {Halperin},\ and\ \citenamefont {Renard}}]{Hagiwara1990}%
  \BibitemOpen
  \bibfield  {author} {\bibinfo {author} {\bibfnamefont {M.}~\bibnamefont
  {Hagiwara}}, \bibinfo {author} {\bibfnamefont {K.}~\bibnamefont {Katsumata}},
  \bibinfo {author} {\bibfnamefont {I.}~\bibnamefont {Affleck}}, \bibinfo
  {author} {\bibfnamefont {B.~I.}\ \bibnamefont {Halperin}}, \ and\ \bibinfo
  {author} {\bibfnamefont {J.~P.}\ \bibnamefont {Renard}},\ }\href@noop {}
  {\bibfield  {journal} {\bibinfo  {journal} {Phys. Rev. Lett.}\ }\textbf
  {\bibinfo {volume} {65}},\ \bibinfo {pages} {3181} (\bibinfo {year}
  {1990})}\BibitemShut {NoStop}%
\bibitem [{\citenamefont {Glarum}\ \emph {et~al.}(1991)\citenamefont {Glarum},
  \citenamefont {Geschwind}, \citenamefont {Lee}, \citenamefont {Kaplan},\ and\
  \citenamefont {Michel}}]{Glarum1991}%
  \BibitemOpen
  \bibfield  {author} {\bibinfo {author} {\bibfnamefont {S.~H.}\ \bibnamefont
  {Glarum}}, \bibinfo {author} {\bibfnamefont {S.}~\bibnamefont {Geschwind}},
  \bibinfo {author} {\bibfnamefont {K.~M.}\ \bibnamefont {Lee}}, \bibinfo
  {author} {\bibfnamefont {M.~L.}\ \bibnamefont {Kaplan}}, \ and\ \bibinfo
  {author} {\bibfnamefont {J.}~\bibnamefont {Michel}},\ }\href@noop {}
  {\bibfield  {journal} {\bibinfo  {journal} {Phys. Rev. Lett.}\ }\textbf
  {\bibinfo {volume} {67}},\ \bibinfo {pages} {1614} (\bibinfo {year}
  {1991})}\BibitemShut {NoStop}%
\bibitem [{\citenamefont {Ma}\ \emph {et~al.}(1992)\citenamefont {Ma},
  \citenamefont {Broholm}, \citenamefont {Reich}, \citenamefont {Sternlieb},\
  and\ \citenamefont {Erwin}}]{Ma1992}%
  \BibitemOpen
  \bibfield  {author} {\bibinfo {author} {\bibfnamefont {S.}~\bibnamefont
  {Ma}}, \bibinfo {author} {\bibfnamefont {C.}~\bibnamefont {Broholm}},
  \bibinfo {author} {\bibfnamefont {D.~H.}\ \bibnamefont {Reich}}, \bibinfo
  {author} {\bibfnamefont {B.~J.}\ \bibnamefont {Sternlieb}}, \ and\ \bibinfo
  {author} {\bibfnamefont {R.~W.}\ \bibnamefont {Erwin}},\ }\href@noop {}
  {\bibfield  {journal} {\bibinfo  {journal} {Phys. Rev. Lett.}\ }\textbf
  {\bibinfo {volume} {69}},\ \bibinfo {pages} {3571} (\bibinfo {year}
  {1992})}\BibitemShut {NoStop}%
\bibitem [{\citenamefont {Zaliznyak}\ \emph {et~al.}(1994)\citenamefont
  {Zaliznyak}, \citenamefont {Regnault},\ and\ \citenamefont
  {Petitgrand}}]{Zaliznyak1994}%
  \BibitemOpen
  \bibfield  {author} {\bibinfo {author} {\bibfnamefont {I.~A.}\ \bibnamefont
  {Zaliznyak}}, \bibinfo {author} {\bibfnamefont {L.-P.}\ \bibnamefont
  {Regnault}}, \ and\ \bibinfo {author} {\bibfnamefont {D.}~\bibnamefont
  {Petitgrand}},\ }\href@noop {} {\bibfield  {journal} {\bibinfo  {journal}
  {Phys. Rev. B}\ }\textbf {\bibinfo {volume} {50}},\ \bibinfo {pages} {15824}
  (\bibinfo {year} {1994})}\BibitemShut {NoStop}%
\bibitem [{\citenamefont {Kenzelmann}\ \emph {et~al.}(2001)\citenamefont
  {Kenzelmann}, \citenamefont {Cowley}, \citenamefont {Buyers},\ and\
  \citenamefont {McMorrow}}]{Kenzelmann2001}%
  \BibitemOpen
  \bibfield  {author} {\bibinfo {author} {\bibfnamefont {M.}~\bibnamefont
  {Kenzelmann}}, \bibinfo {author} {\bibfnamefont {R.~A.}\ \bibnamefont
  {Cowley}}, \bibinfo {author} {\bibfnamefont {W.~J.~L.}\ \bibnamefont
  {Buyers}}, \ and\ \bibinfo {author} {\bibfnamefont {D.~F.}\ \bibnamefont
  {McMorrow}},\ }\href@noop {} {\bibfield  {journal} {\bibinfo  {journal}
  {Phys. Rev. B}\ }\textbf {\bibinfo {volume} {63}},\ \bibinfo {pages} {134417}
  (\bibinfo {year} {2001})}\BibitemShut {NoStop}%
\bibitem [{\citenamefont {Kenzelmann}\ \emph {et~al.}(2002)\citenamefont
  {Kenzelmann}, \citenamefont {Cowley}, \citenamefont {Buyers}, \citenamefont
  {Tun}, \citenamefont {Coldea},\ and\ \citenamefont
  {Enderle}}]{Kenzelmann2002}%
  \BibitemOpen
  \bibfield  {author} {\bibinfo {author} {\bibfnamefont {M.}~\bibnamefont
  {Kenzelmann}}, \bibinfo {author} {\bibfnamefont {R.~A.}\ \bibnamefont
  {Cowley}}, \bibinfo {author} {\bibfnamefont {W.~J.~L.}\ \bibnamefont
  {Buyers}}, \bibinfo {author} {\bibfnamefont {Z.}~\bibnamefont {Tun}},
  \bibinfo {author} {\bibfnamefont {R.}~\bibnamefont {Coldea}}, \ and\ \bibinfo
  {author} {\bibfnamefont {M.}~\bibnamefont {Enderle}},\ }\href@noop {}
  {\bibfield  {journal} {\bibinfo  {journal} {Phys. Rev. B}\ }\textbf {\bibinfo
  {volume} {66}},\ \bibinfo {pages} {024407} (\bibinfo {year}
  {2002})}\BibitemShut {NoStop}%
\bibitem [{\citenamefont {Masuda}\ \emph {et~al.}(2002)\citenamefont {Masuda},
  \citenamefont {Sakaguchi},\ and\ \citenamefont {Uchinokura}}]{Masuda2002}%
  \BibitemOpen
  \bibfield  {author} {\bibinfo {author} {\bibfnamefont {T.}~\bibnamefont
  {Masuda}}, \bibinfo {author} {\bibfnamefont {T.}~\bibnamefont {Sakaguchi}}, \
  and\ \bibinfo {author} {\bibfnamefont {K.}~\bibnamefont {Uchinokura}},\
  }\href@noop {} {\bibfield  {journal} {\bibinfo  {journal} {J. Phys. Soc.
  Jpn.}\ }\textbf {\bibinfo {volume} {71}},\ \bibinfo {pages} {2637} (\bibinfo
  {year} {2002})}\BibitemShut {NoStop}%
\bibitem [{\citenamefont {Uchiyama}\ \emph {et~al.}(1999)\citenamefont
  {Uchiyama}, \citenamefont {Sasago}, \citenamefont {Tsukada}, \citenamefont
  {Uchinokura}, \citenamefont {Zheludev}, \citenamefont {Hayashi},
  \citenamefont {Miura},\ and\ \citenamefont {B{\"{o}}ni}}]{Uchiyama1999}%
  \BibitemOpen
  \bibfield  {author} {\bibinfo {author} {\bibfnamefont {Y.}~\bibnamefont
  {Uchiyama}}, \bibinfo {author} {\bibfnamefont {Y.}~\bibnamefont {Sasago}},
  \bibinfo {author} {\bibfnamefont {I.}~\bibnamefont {Tsukada}}, \bibinfo
  {author} {\bibfnamefont {K.}~\bibnamefont {Uchinokura}}, \bibinfo {author}
  {\bibfnamefont {A.}~\bibnamefont {Zheludev}}, \bibinfo {author}
  {\bibfnamefont {T.}~\bibnamefont {Hayashi}}, \bibinfo {author} {\bibfnamefont
  {N.}~\bibnamefont {Miura}}, \ and\ \bibinfo {author} {\bibfnamefont
  {P.}~\bibnamefont {B{\"{o}}ni}},\ }\href@noop {} {\bibfield  {journal}
  {\bibinfo  {journal} {Phys. Rev. Lett.}\ }\textbf {\bibinfo {volume} {83}},\
  \bibinfo {pages} {632} (\bibinfo {year} {1999})}\BibitemShut {NoStop}%
\bibitem [{\citenamefont {Pahari}\ \emph {et~al.}(2006)\citenamefont {Pahari},
  \citenamefont {Ghoshray}, \citenamefont {Sarkar}, \citenamefont
  {Bandyopadhyay},\ and\ \citenamefont {Ghoshray}}]{Pahari2006}%
  \BibitemOpen
  \bibfield  {author} {\bibinfo {author} {\bibfnamefont {B.}~\bibnamefont
  {Pahari}}, \bibinfo {author} {\bibfnamefont {K.}~\bibnamefont {Ghoshray}},
  \bibinfo {author} {\bibfnamefont {R.}~\bibnamefont {Sarkar}}, \bibinfo
  {author} {\bibfnamefont {B.}~\bibnamefont {Bandyopadhyay}}, \ and\ \bibinfo
  {author} {\bibfnamefont {A.}~\bibnamefont {Ghoshray}},\ }\href@noop {}
  {\bibfield  {journal} {\bibinfo  {journal} {Phys. Rev. B}\ }\textbf {\bibinfo
  {volume} {73}},\ \bibinfo {pages} {012407} (\bibinfo {year}
  {2006})}\BibitemShut {NoStop}%
\bibitem [{\citenamefont {He}\ and\ \citenamefont {Ueda}(2008)}]{He2008}%
  \BibitemOpen
  \bibfield  {author} {\bibinfo {author} {\bibfnamefont {Z.}~\bibnamefont
  {He}}\ and\ \bibinfo {author} {\bibfnamefont {Y.}~\bibnamefont {Ueda}},\
  }\href@noop {} {\bibfield  {journal} {\bibinfo  {journal} {J. Phys. Soc.
  Jpn.}\ }\textbf {\bibinfo {volume} {77}},\ \bibinfo {pages} {013703}
  (\bibinfo {year} {2008})}\BibitemShut {NoStop}%
\bibitem [{\citenamefont {Hikihara}\ \emph {et~al.}(2001)\citenamefont
  {Hikihara}, \citenamefont {Kaburagi},\ and\ \citenamefont
  {Kawamura}}]{Hikihara2001}%
  \BibitemOpen
  \bibfield  {author} {\bibinfo {author} {\bibfnamefont {T.}~\bibnamefont
  {Hikihara}}, \bibinfo {author} {\bibfnamefont {M.}~\bibnamefont {Kaburagi}},
  \ and\ \bibinfo {author} {\bibfnamefont {H.}~\bibnamefont {Kawamura}},\
  }\href@noop {} {\bibfield  {journal} {\bibinfo  {journal} {Phys. Rev. B}\
  }\textbf {\bibinfo {volume} {63}},\ \bibinfo {pages} {174430} (\bibinfo
  {year} {2001})}\BibitemShut {NoStop}%
\bibitem [{\citenamefont {Kolezhuk}\ \emph {et~al.}(1996)\citenamefont
  {Kolezhuk}, \citenamefont {Roth},\ and\ \citenamefont
  {Schollw{\"{o}}ck}}]{Kolezhuk1996}%
  \BibitemOpen
  \bibfield  {author} {\bibinfo {author} {\bibfnamefont {A.}~\bibnamefont
  {Kolezhuk}}, \bibinfo {author} {\bibfnamefont {R.}~\bibnamefont {Roth}}, \
  and\ \bibinfo {author} {\bibfnamefont {U.}~\bibnamefont {Schollw{\"{o}}ck}},\
  }\href@noop {} {\bibfield  {journal} {\bibinfo  {journal} {Phys. Rev. Lett.}\
  }\textbf {\bibinfo {volume} {77}},\ \bibinfo {pages} {5142} (\bibinfo {year}
  {1996})}\BibitemShut {NoStop}%
\bibitem [{\citenamefont {Kolezhuk}\ \emph {et~al.}(1997)\citenamefont
  {Kolezhuk}, \citenamefont {Roth},\ and\ \citenamefont
  {Schollw\"ock}}]{Kolezhuk1997}%
  \BibitemOpen
  \bibfield  {author} {\bibinfo {author} {\bibfnamefont {A.}~\bibnamefont
  {Kolezhuk}}, \bibinfo {author} {\bibfnamefont {R.}~\bibnamefont {Roth}}, \
  and\ \bibinfo {author} {\bibfnamefont {U.}~\bibnamefont {Schollw\"ock}},\
  }\href@noop {} {\bibfield  {journal} {\bibinfo  {journal} {Phys. Rev. B}\
  }\textbf {\bibinfo {volume} {55}},\ \bibinfo {pages} {8928} (\bibinfo {year}
  {1997})}\BibitemShut {NoStop}%
\bibitem [{\citenamefont {Pixley}\ \emph {et~al.}(2014)\citenamefont {Pixley},
  \citenamefont {Shashi},\ and\ \citenamefont {Nevidomskyy}}]{Pixley2014}%
  \BibitemOpen
  \bibfield  {author} {\bibinfo {author} {\bibfnamefont {J.~H.}\ \bibnamefont
  {Pixley}}, \bibinfo {author} {\bibfnamefont {A.}~\bibnamefont {Shashi}}, \
  and\ \bibinfo {author} {\bibfnamefont {A.~H.}\ \bibnamefont {Nevidomskyy}},\
  }\href@noop {} {\bibfield  {journal} {\bibinfo  {journal} {Phys. Rev. B}\
  }\textbf {\bibinfo {volume} {90}},\ \bibinfo {pages} {214426} (\bibinfo
  {year} {2014})}\BibitemShut {NoStop}%
\bibitem [{\citenamefont {Hamada}\ \emph {et~al.}(1988)\citenamefont {Hamada},
  \citenamefont {Kane}, \citenamefont {Nakagawa},\ and\ \citenamefont
  {Natsume}}]{Hamada1988}%
  \BibitemOpen
  \bibfield  {author} {\bibinfo {author} {\bibfnamefont {T.}~\bibnamefont
  {Hamada}}, \bibinfo {author} {\bibfnamefont {J.}~\bibnamefont {Kane}},
  \bibinfo {author} {\bibfnamefont {S.}~\bibnamefont {Nakagawa}}, \ and\
  \bibinfo {author} {\bibfnamefont {Y.}~\bibnamefont {Natsume}},\ }\href@noop
  {} {\bibfield  {journal} {\bibinfo  {journal} {J. Phys. Soc. Jpn.}\ }\textbf
  {\bibinfo {volume} {57}},\ \bibinfo {pages} {1891} (\bibinfo {year}
  {1988})}\BibitemShut {NoStop}%
\bibitem [{\citenamefont {Allen}\ and\ \citenamefont
  {S{\'{e}}n{\'{e}}chal}(1997)}]{Allen1997}%
  \BibitemOpen
  \bibfield  {author} {\bibinfo {author} {\bibfnamefont {D.}~\bibnamefont
  {Allen}}\ and\ \bibinfo {author} {\bibfnamefont {D.}~\bibnamefont
  {S{\'{e}}n{\'{e}}chal}},\ }\href@noop {} {\bibfield  {journal} {\bibinfo
  {journal} {Phys. Rev. B}\ }\textbf {\bibinfo {volume} {55}},\ \bibinfo
  {pages} {299} (\bibinfo {year} {1997})}\BibitemShut {NoStop}%
\bibitem [{\citenamefont {H{\"{a}}rtel}\ \emph {et~al.}(2008)\citenamefont
  {H{\"{a}}rtel}, \citenamefont {Richter}, \citenamefont {Ihle},\ and\
  \citenamefont {Drechsler}}]{Hartel2008}%
  \BibitemOpen
  \bibfield  {author} {\bibinfo {author} {\bibfnamefont {M.}~\bibnamefont
  {H{\"{a}}rtel}}, \bibinfo {author} {\bibfnamefont {J.}~\bibnamefont
  {Richter}}, \bibinfo {author} {\bibfnamefont {D.}~\bibnamefont {Ihle}}, \
  and\ \bibinfo {author} {\bibfnamefont {S.-L.}\ \bibnamefont {Drechsler}},\
  }\href@noop {} {\bibfield  {journal} {\bibinfo  {journal} {Phys. Rev. B}\
  }\textbf {\bibinfo {volume} {78}},\ \bibinfo {pages} {174412} (\bibinfo
  {year} {2008})}\BibitemShut {NoStop}%
\bibitem [{\citenamefont {Sudan}\ \emph {et~al.}(2009)\citenamefont {Sudan},
  \citenamefont {L{\"{u}}scher},\ and\ \citenamefont
  {L{\"{a}}uchli}}]{Sudan2009}%
  \BibitemOpen
  \bibfield  {author} {\bibinfo {author} {\bibfnamefont {J.}~\bibnamefont
  {Sudan}}, \bibinfo {author} {\bibfnamefont {A.}~\bibnamefont
  {L{\"{u}}scher}}, \ and\ \bibinfo {author} {\bibfnamefont {A.~M.}\
  \bibnamefont {L{\"{a}}uchli}},\ }\href@noop {} {\bibfield  {journal}
  {\bibinfo  {journal} {Phys. Rev. B}\ }\textbf {\bibinfo {volume} {80}},\
  \bibinfo {pages} {140402(R)} (\bibinfo {year} {2009})}\BibitemShut {NoStop}%
\bibitem [{\citenamefont {Sirker}(2010)}]{Sirker2010}%
  \BibitemOpen
  \bibfield  {author} {\bibinfo {author} {\bibfnamefont {J.}~\bibnamefont
  {Sirker}},\ }\href@noop {} {\bibfield  {journal} {\bibinfo  {journal} {Phys.
  Rev. B}\ }\textbf {\bibinfo {volume} {81}},\ \bibinfo {pages} {014419}
  (\bibinfo {year} {2010})}\BibitemShut {NoStop}%
\bibitem [{\citenamefont {Sirker}\ \emph {et~al.}(2011)\citenamefont {Sirker},
  \citenamefont {Krivnov}, \citenamefont {Dmitriev}, \citenamefont {Herzog},
  \citenamefont {Janson}, \citenamefont {Nishimoto}, \citenamefont
  {Drechsler},\ and\ \citenamefont {Richter}}]{Sirker2011}%
  \BibitemOpen
  \bibfield  {author} {\bibinfo {author} {\bibfnamefont {J.}~\bibnamefont
  {Sirker}}, \bibinfo {author} {\bibfnamefont {V.~Y.}\ \bibnamefont {Krivnov}},
  \bibinfo {author} {\bibfnamefont {D.~V.}\ \bibnamefont {Dmitriev}}, \bibinfo
  {author} {\bibfnamefont {A.}~\bibnamefont {Herzog}}, \bibinfo {author}
  {\bibfnamefont {O.}~\bibnamefont {Janson}}, \bibinfo {author} {\bibfnamefont
  {S.}~\bibnamefont {Nishimoto}}, \bibinfo {author} {\bibfnamefont {S.~L.}\
  \bibnamefont {Drechsler}}, \ and\ \bibinfo {author} {\bibfnamefont
  {J.}~\bibnamefont {Richter}},\ }\href@noop {} {\bibfield  {journal} {\bibinfo
   {journal} {Phys. Rev. B}\ }\textbf {\bibinfo {volume} {84}},\ \bibinfo
  {pages} {144403} (\bibinfo {year} {2011})}\BibitemShut {NoStop}%
\bibitem [{\citenamefont {Furukawa}\ \emph {et~al.}(2010)\citenamefont
  {Furukawa}, \citenamefont {Sato},\ and\ \citenamefont
  {Onoda}}]{Furukawa2010}%
  \BibitemOpen
  \bibfield  {author} {\bibinfo {author} {\bibfnamefont {S.}~\bibnamefont
  {Furukawa}}, \bibinfo {author} {\bibfnamefont {M.}~\bibnamefont {Sato}}, \
  and\ \bibinfo {author} {\bibfnamefont {S.}~\bibnamefont {Onoda}},\
  }\href@noop {} {\bibfield  {journal} {\bibinfo  {journal} {Phys. Rev. Lett.}\
  }\textbf {\bibinfo {volume} {105}},\ \bibinfo {pages} {257205} (\bibinfo
  {year} {2010})}\BibitemShut {NoStop}%
\bibitem [{\citenamefont {Furukawa}\ \emph {et~al.}(2012)\citenamefont
  {Furukawa}, \citenamefont {Sato}, \citenamefont {Onoda},\ and\ \citenamefont
  {Furusaki}}]{Furukawa2012}%
  \BibitemOpen
  \bibfield  {author} {\bibinfo {author} {\bibfnamefont {S.}~\bibnamefont
  {Furukawa}}, \bibinfo {author} {\bibfnamefont {M.}~\bibnamefont {Sato}},
  \bibinfo {author} {\bibfnamefont {S.}~\bibnamefont {Onoda}}, \ and\ \bibinfo
  {author} {\bibfnamefont {A.}~\bibnamefont {Furusaki}},\ }\href@noop {}
  {\bibfield  {journal} {\bibinfo  {journal} {Phys. Rev. B}\ }\textbf {\bibinfo
  {volume} {86}},\ \bibinfo {pages} {094417} (\bibinfo {year}
  {2012})}\BibitemShut {NoStop}%
\bibitem [{\citenamefont {Kennedy}\ and\ \citenamefont
  {Tasaki}(1992)}]{Kennedy1992}%
  \BibitemOpen
  \bibfield  {author} {\bibinfo {author} {\bibfnamefont {T.}~\bibnamefont
  {Kennedy}}\ and\ \bibinfo {author} {\bibfnamefont {H.}~\bibnamefont
  {Tasaki}},\ }\href@noop {} {\bibfield  {journal} {\bibinfo  {journal} {Phys.
  Rev. B}\ }\textbf {\bibinfo {volume} {45}},\ \bibinfo {pages} {304} (\bibinfo
  {year} {1992})}\BibitemShut {NoStop}%
\bibitem [{\citenamefont {Nomura}\ and\ \citenamefont
  {Takada}(1991)}]{Nomura1991}%
  \BibitemOpen
  \bibfield  {author} {\bibinfo {author} {\bibfnamefont {K.}~\bibnamefont
  {Nomura}}\ and\ \bibinfo {author} {\bibfnamefont {S.}~\bibnamefont
  {Takada}},\ }\href@noop {} {\bibfield  {journal} {\bibinfo  {journal} {J.
  Phys. Soc. Jpn.}\ }\textbf {\bibinfo {volume} {60}},\ \bibinfo {pages} {389}
  (\bibinfo {year} {1991})}\BibitemShut {NoStop}%
\bibitem [{\citenamefont {Schollw{\"{o}}ck}(2005)}]{Schollwock2005}%
  \BibitemOpen
  \bibfield  {author} {\bibinfo {author} {\bibfnamefont {U.}~\bibnamefont
  {Schollw{\"{o}}ck}},\ }\href@noop {} {\bibfield  {journal} {\bibinfo
  {journal} {Rev. Mod. Phys.}\ }\textbf {\bibinfo {volume} {77}},\ \bibinfo
  {pages} {259} (\bibinfo {year} {2005})}\BibitemShut {NoStop}%
\bibitem [{\citenamefont {Bader}\ and\ \citenamefont
  {Schilling}(1979)}]{Bader1979}%
  \BibitemOpen
  \bibfield  {author} {\bibinfo {author} {\bibfnamefont {H.~P.}\ \bibnamefont
  {Bader}}\ and\ \bibinfo {author} {\bibfnamefont {R.}~\bibnamefont
  {Schilling}},\ }\href@noop {} {\bibfield  {journal} {\bibinfo  {journal}
  {Phys. Rev. B}\ }\textbf {\bibinfo {volume} {19}},\ \bibinfo {pages} {3556}
  (\bibinfo {year} {1979})}\BibitemShut {NoStop}%
\bibitem [{\citenamefont {den Nijs}\ and\ \citenamefont
  {Rommelse}(1989)}]{DenNijs1989}%
  \BibitemOpen
  \bibfield  {author} {\bibinfo {author} {\bibfnamefont {M.}~\bibnamefont {den
  Nijs}}\ and\ \bibinfo {author} {\bibfnamefont {K.}~\bibnamefont {Rommelse}},\
  }\href@noop {} {\bibfield  {journal} {\bibinfo  {journal} {Phys. Rev. B}\
  }\textbf {\bibinfo {volume} {40}},\ \bibinfo {pages} {4709} (\bibinfo {year}
  {1989})}\BibitemShut {NoStop}%
\bibitem [{\citenamefont {Girvin}\ and\ \citenamefont
  {Arovas}(1989)}]{Girvin1989}%
  \BibitemOpen
  \bibfield  {author} {\bibinfo {author} {\bibfnamefont {S.~M.}\ \bibnamefont
  {Girvin}}\ and\ \bibinfo {author} {\bibfnamefont {D.~P.}\ \bibnamefont
  {Arovas}},\ }\href@noop {} {\bibfield  {journal} {\bibinfo  {journal} {Phys.
  Scr.}\ }\textbf {\bibinfo {volume} {T27}},\ \bibinfo {pages} {156} (\bibinfo
  {year} {1989})}\BibitemShut {NoStop}%
\bibitem [{\citenamefont {Kolezhuk}\ and\ \citenamefont
  {Schollw\"ock}(2002)}]{Kolezhuk2002}%
  \BibitemOpen
  \bibfield  {author} {\bibinfo {author} {\bibfnamefont {A.~K.}\ \bibnamefont
  {Kolezhuk}}\ and\ \bibinfo {author} {\bibfnamefont {U.}~\bibnamefont
  {Schollw\"ock}},\ }\href@noop {} {\bibfield  {journal} {\bibinfo  {journal}
  {Phys. Rev. B}\ }\textbf {\bibinfo {volume} {65}},\ \bibinfo {pages} {100401}
  (\bibinfo {year} {2002})}\BibitemShut {NoStop}%
\bibitem [{\citenamefont {Balents}\ and\ \citenamefont
  {Starykh}(2016)}]{Balents2016}%
  \BibitemOpen
  \bibfield  {author} {\bibinfo {author} {\bibfnamefont {L.}~\bibnamefont
  {Balents}}\ and\ \bibinfo {author} {\bibfnamefont {O.~A.}\ \bibnamefont
  {Starykh}},\ }\href@noop {} {\bibfield  {journal} {\bibinfo  {journal} {Phys.
  Rev. Lett.}\ }\textbf {\bibinfo {volume} {116}},\ \bibinfo {pages} {177201}
  (\bibinfo {year} {2016})}\BibitemShut {NoStop}%
\bibitem [{\citenamefont {Senko}\ \emph {et~al.}(2015)\citenamefont {Senko},
  \citenamefont {Richerme}, \citenamefont {Smith}, \citenamefont {Lee},
  \citenamefont {Cohen}, \citenamefont {Retzker},\ and\ \citenamefont
  {Monroe}}]{Senko2015}%
  \BibitemOpen
  \bibfield  {author} {\bibinfo {author} {\bibfnamefont {C.}~\bibnamefont
  {Senko}}, \bibinfo {author} {\bibfnamefont {P.}~\bibnamefont {Richerme}},
  \bibinfo {author} {\bibfnamefont {J.}~\bibnamefont {Smith}}, \bibinfo
  {author} {\bibfnamefont {A.}~\bibnamefont {Lee}}, \bibinfo {author}
  {\bibfnamefont {I.}~\bibnamefont {Cohen}}, \bibinfo {author} {\bibfnamefont
  {A.}~\bibnamefont {Retzker}}, \ and\ \bibinfo {author} {\bibfnamefont
  {C.}~\bibnamefont {Monroe}},\ }\href@noop {} {\bibfield  {journal} {\bibinfo
  {journal} {Phys. Rev. X}\ }\textbf {\bibinfo {volume} {5}},\ \bibinfo {pages}
  {021026} (\bibinfo {year} {2015})}\BibitemShut {NoStop}%
\bibitem [{\citenamefont {Cohen}\ \emph {et~al.}(2015)\citenamefont {Cohen},
  \citenamefont {Richerme}, \citenamefont {Gong}, \citenamefont {Monroe},\ and\
  \citenamefont {Retzker}}]{Cohen2015}%
  \BibitemOpen
  \bibfield  {author} {\bibinfo {author} {\bibfnamefont {I.}~\bibnamefont
  {Cohen}}, \bibinfo {author} {\bibfnamefont {P.}~\bibnamefont {Richerme}},
  \bibinfo {author} {\bibfnamefont {Z.-X.}\ \bibnamefont {Gong}}, \bibinfo
  {author} {\bibfnamefont {C.}~\bibnamefont {Monroe}}, \ and\ \bibinfo {author}
  {\bibfnamefont {A.}~\bibnamefont {Retzker}},\ }\href@noop {} {\bibfield
  {journal} {\bibinfo  {journal} {Phys. Rev. A}\ }\textbf {\bibinfo {volume}
  {92}},\ \bibinfo {pages} {012334} (\bibinfo {year} {2015})}\BibitemShut
  {NoStop}%
\bibitem [{\citenamefont {Garc{\'{i}}a-Ripoll}\ \emph
  {et~al.}(2004)\citenamefont {Garc{\'{i}}a-Ripoll}, \citenamefont
  {Martin-Delgado},\ and\ \citenamefont {Cirac}}]{Garcia-Ripoll2004}%
  \BibitemOpen
  \bibfield  {author} {\bibinfo {author} {\bibfnamefont {J.~J.}\ \bibnamefont
  {Garc{\'{i}}a-Ripoll}}, \bibinfo {author} {\bibfnamefont {M.~A.}\
  \bibnamefont {Martin-Delgado}}, \ and\ \bibinfo {author} {\bibfnamefont
  {J.~I.}\ \bibnamefont {Cirac}},\ }\href@noop {} {\bibfield  {journal}
  {\bibinfo  {journal} {Phys. Rev. Lett.}\ }\textbf {\bibinfo {volume} {93}},\
  \bibinfo {pages} {250405} (\bibinfo {year} {2004})}\BibitemShut {NoStop}%
\bibitem [{\citenamefont {Hikihara}(2002)}]{Hikihara2002}%
  \BibitemOpen
  \bibfield  {author} {\bibinfo {author} {\bibfnamefont {T.}~\bibnamefont
  {Hikihara}},\ }\href@noop {} {\bibfield  {journal} {\bibinfo  {journal} {J.
  Phys. Soc. Jpn.}\ }\textbf {\bibinfo {volume} {71}},\ \bibinfo {pages} {319}
  (\bibinfo {year} {2002})}\BibitemShut {NoStop}%
\end{thebibliography}%

\end{document}